\documentclass[twocolumn]{aastex631}

\graphicspath{{./}{figures/}}
\usepackage{gensymb}
\usepackage{amsmath}
\usepackage{appendix}
\usepackage{flafter}
\usepackage{xspace}
\usepackage{tablefootnote}
\usepackage[flushleft]{threeparttable}
\usepackage{graphicx}

\newcommand{\Ni}{\emph{NICER}\xspace}
\newcommand{\sw}{\emph{Swift}\xspace}
\newcommand{\xmm}{\emph{XMM-Newton}\xspace}
\newcommand{\cha}{\emph{Chandra}\xspace}

\newcommand{\avd}{AT~2019avd\xspace}
\newcommand{\ergs}{erg\,s$^{-1}$\xspace}
\newcommand{\kms}{km\,s$^{-1}$\xspace}

\shorttitle{Rapid dimming in \avd}
\shortauthors{Wang et al.}

\begin{document}

\title{\centering Rapid dimming followed by a state transition: a study of the highly variable nuclear transient {\avd} over 1000+\,days}

\author{Yanan Wang}
\affiliation{National Astronomical Observatories, Chinese Academy of Sciences, 20A Datun Road, Beijing 100101, China}
\affiliation{Physics \& Astronomy, University of Southampton, Southampton, Hampshire SO17~1BJ, UK}

\author{Dheeraj R. Pasham}
\affiliation{Kavli Institute for Astrophysics and Space Research, Massachusetts Institute of Technology, Cambridge MA 02139, USA}

\author{Diego Altamirano}
\affiliation{Physics \& Astronomy, University of Southampton, Southampton, Hampshire SO17~1BJ, UK}

\author{Andr{\'e}s G{\'u}rpide}
\affiliation{Physics \& Astronomy, University of Southampton, Southampton, Hampshire SO17~1BJ, UK}

\author{Noel Castro Segura}
\affiliation{Physics \& Astronomy, University of Southampton, Southampton, Hampshire SO17~1BJ, UK}

\author{Matthew Middleton}
\affiliation{Physics \& Astronomy, University of Southampton, Southampton, Hampshire SO17~1BJ, UK}

\author{Long Ji}
\affiliation{School of Physics and Astronomy, Sun Yat-sen University, 2 Daxue Road, Zhuhai, Guangdong 519082, China}

\author{Santiago del Palacio}
\affiliation{Department of Space, Earth and Environment, Chalmers University of Technology, SE-412 96 Gothenburg, Sweden}

\author{Muryel Guolo}
\affiliation{Department of Physics and Astronomy, Johns Hopkins University, 3400 N. Charles St., Baltimore MD 21218, USA}

\author{Poshak Gandhi}
\affiliation{Physics \& Astronomy, University of Southampton, Southampton, Hampshire SO17~1BJ, UK}

\author{Shuang-Nan Zhang}
\affiliation{Key Laboratory for Particle Astrophysics, Institute of High Energy Physics, Chinese Academy of Sciences, 19B Yuquan Road, Beijing 100049, China}
\affiliation{University of Chinese Academy of Sciences, Chinese Academy of Sciences, Beijing 100049, China}

\author{Ronald Remillard}
\affiliation{Kavli Institute for Astrophysics and Space Research, Massachusetts Institute of Technology, Cambridge MA 02139, USA}

\author{Dacheng Lin}
\affiliation{Department of Physics, Northeastern University, Boston, MA 02115-5000, USA}

\author{Megan Masterson}
\affiliation{Kavli Institute for Astrophysics and Space Research, Massachusetts Institute of Technology, Cambridge MA 02139, USA}

\author{Ranieri D. Baldi}
\affiliation{INAF - Istituto di Radioastronomia, Via P. Gobetti 101, I-40129 Bologna, Italy}
\affiliation{Physics \& Astronomy, University of Southampton, Southampton, Hampshire SO17~1BJ, UK}

\author{Francesco Tombesi}
\affiliation{Dipartimento di Fisica, Univerisit{\`a} di Roma Tor Vergata, via della Ricerca Scientifica 1, I-00133 Roma, Italy}
\affiliation{INAF – Osservatorio Astronomico di Roma, Via Frascati 33, 00040 Monte Porzio Catone, Italy}
\affiliation{Department of Astronomy, University of Maryland, College Park, MD 20742, USA}
\affiliation{NASA/Goddard Space Flight Center, Code 662, Greenbelt, MD 20771, USA}

\author{Jon M. Miller}
\affiliation{Department of Astronomy, University of Michigan, 1085 South University Avenue, Ann Arbor, MI, 48109, USA}

\author{Wenda Zhang}
\affiliation{National Astronomical Observatories, Chinese Academy of Sciences, 20A Datun Road, Beijing 100101, China}

\author{Andrea Sanna}
\affiliation{Dipartimento di Fisica, Universit{\`a} degli Studi di Cagliari, SP Monserrato-Sestu km 0.7, 09042 Monserrato, Italy}

\begin{abstract}
The tidal disruption of a star around a supermassive black hole (SMBH) offers a unique opportunity to study accretion onto a SMBH on a human-timescale. We present results from our 1000+\,days \Ni, \sw and \cha monitoring campaign of \avd, a nuclear transient with TDE-like properties. Our primary finding is that approximately 225\,days following the peak of X-ray emission, there is a rapid drop in luminosity exceeding two orders of magnitude. 
This X-ray drop-off is accompanied by X-ray spectral hardening, followed by a 740-day plateau phase. During this phase, the spectral index decreases from $6.2\pm1.1$ to $2.3\pm0.4$, while the disk temperature remains constant. 
Additionally, we detect pronounced X-ray variability, with an average fractional root mean squared amplitude of 47\%, manifesting over timescales of a few dozen minutes. We propose that this phenomenon may be attributed to intervening clumpy outflows. 
The overall properties of \avd suggest that the accretion disk evolves from a super-Eddington to a sub-Eddington luminosity state, possibly associated with a compact jet. This evolution follows a pattern in the hardness-intensity diagram similar to that observed in stellar-mass black holes, supporting the mass invariance of accretion-ejection processes around black holes.

\end{abstract}


\section{Introduction} \label{sec:intro}
Accretion processes have been studied in systems ranging from stellar-mass to supermassive black holes (SMBHs; masses $>$10$^{5}$~$M_{\odot}$), spanning up to eight orders of magnitude in mass (e.g. \citealt{Rees1984,Fender2004,Remillard2006,King2015,Gezari2021}). 
At the lower end of the mass range and at sub-Eddington accretion rates, we find black hole X-ray binaries (BHXRBs). These systems are mostly transients that are only observable when they are in a short-lived outburst, typically lasting for a few months to a few years (e.g. \citealt{White1984,Hjellming1999,Homan2003,Belloni2005}). Two main spectral states have been defined based on the co-evolution of two X-ray spectral components, an accretion disk and a corona. In such systems, a hard state is defined when the non-thermal/Comptonization emission from a corona dominates the spectrum, while a soft state is defined when the thermal disk emission dominates the spectrum (see \citealt{Remillard2006} for a review). 
Moreover, high variability represented by a fractional rms of $\sim 20$--$40\%$ has been commonly observed in X-rays in the hard state, while it decreases to $\lesssim 5\%$ in the soft state \citep{Gleissner2004,Munoz2011}. State transitions occur as the dominant component evolves from the Comptonization to the thermal component or vice versa. 

As the mass accretion rate further increases, the system enters the so-called super-Eddington regime, in which intense radiation pressure is expected to drive powerful winds off the disk (e.g. \citealt{Middleton2015b,Pinto2016}), which has also been demonstrated by theoretical works (e.g. \citealt{Shakura1973,Lipunova1999}) and numerical simulations (e.g. \citealt{Narayan2017}). 
Most ultraluminous X-ray sources (ULXs, \citealt{Kaaret2017}) are thought to be accreting in such regime, making them ideal laboratories to study sustained super-Eddington accretion, given their close proximity ($\lesssim$17\,Mpc). As opposed to BHXRBs, they are in general persistent systems and show a variable-when-softer behavior \citep{Middleton2011,Sutton2013,Middleton2015}. 
Although the spectral states are defined differently from BHXRBs, state transitions have also been observed in such systems, and have been associated with changes of the mass accretion rate and the opening angle of the super-critical funnel formed by the winds (\citealt{Sutton2013,Middleton2015,Gurpide2021}). 

At the higher end of the mass range, SMBHs have been argued to be a scaled-up version of BHXRBs. However, the long evolutionary timescale (hundreds of thousands of years) of SMBHs hinders making direct observational comparisons. 
Nuclear transients such as tidal disruption events (TDEs), occurring in the vicinity of SMBHs, are key to solving this issue. The evolution of TDEs across the electromagnetic spectrum occurs on observable timescales, i.e. months-to-years, which makes them a perfect target for understanding the impact of black hole mass on accretion processes in both super- and sub-Eddington regimes.

\cite{Hills1975} predicted that a TDE occurs when a star is disrupted by the gravitational tidal forces of a SMBH with a mass of $10^6$--$10^8\,M_\odot$. 
More than 20 years later, such candidates were detected in X-rays with {\it ROSAT} by \cite{Komossa1999}. With the implementation of multi-wavelength wide-field surveys, the discovery pace of TDEs has accelerated dramatically (around 100 candidates as of now); of which, most are optically/UV selected, while $\sim$30\% have shown X-ray emission and around a dozen have been detected in radio (see \citealt{Velzen2020}, \citealt{Saxton2021}, \citealt{Alexander2020} and \citealt{Gezari2021} for reviews).

In the optical band, TDEs are characterized by their extreme variability on long-term (months-to-years) timescales, large peak luminosities (up to $10^{45}$\,\ergs) and complex optical spectral features (e.g. transient H$\alpha$/He~{\sc ii}/Bowen fluorescence emission lines, e.g. \citealt{Gezari2012,Blagorodnova2018,van_Velzen_21}). 
TDEs are sometimes accompanied by ultra-soft X-ray emission. Unlike their smooth evolution in the optical, their behavior in X-rays is more complex and varies from system to system. 
The X-ray spectrum of TDEs can be described by either a blackbody component with $kT_{\rm bb}=40$--$250$\,eV or a steep powerlaw component with $\Gamma>4$ (e.g. \citealt{Komossa1999,Auchettl2017}). 
There are a handful of systems that exhibit additional spectral features, such as \mbox{ASASSN$-$14li} \citep{Miller2015,Kara2018}, Swift~J1644+57 \citep{Kara2016} and AT2021ehb \citep{yao2022}. In \mbox{ASASSN$-$14li}, highly ionized and blueshifted narrow absorption lines have been discovered in its high-resolution spectra \citep{Miller2015}. \cite{Kara2018} has also identified broad ($\sim$30,000\,km s$^{-1}$) features which were interpreted as an ultrafast outflow with a velocity of 0.2c. In Swift~J1644+57, \cite{Kara2016} observed a redshifted iron K$\alpha$ line in the 5.5--8\,keV energy range, which was considered as evidence of disk reflection. 
In AT2018fyk, its spectrum shows both a soft excess and a hard tail in the 0.3--10\,keV; a state transition has also been observed, where the spectrum changes from disc-dominated to powerlaw-dominated when $L_{\rm bol} \sim 0.02\,L_{\rm Edd}$ \citep{Wevers2021}.

The flare of TDEs does not always decline smoothly, but are accompanied by relatively short-term X-ray variability (e.g. \citealt{Saxton2012a,Pasham2022}). Such variability has been observed in timescales of hundreds to thousands of seconds, e.g. mHz quasi-periodic oscillations (QPOs; \citealt{Reis2012,Pasham2019}), sub-mHz time lags \citep{Kara2016,Jin2021}, dips with peculiar patterns in the lightcurve \citep{Saxton2012a}, and variability associated with a softer-when-dimmer behavior \citep{Lin2015}. Interestingly, \cite{Pasham2019} discovered a stable 131-second QPO in ASASSN--14li, whose frequency is comparable to the mHz QPOs observed in BHXRBs (e.g. \citealt{Altamirano2012}) and ULXs (e.g. \citealt{Strohmayer2003,Mucciarelli2006}), but in softer X-rays (0.3--1\,keV). This discovery disfavors the correlation between the QPO and hard X-rays, and the QPO-mass scaling.

A peculiar nuclear transient {\avd}, located at $z=0.028$, has been detected from radio to soft X-rays. This transient was first discovered in the optical by the Zwicky Transient Facility (ZTF; \citealt{Bellm2019}) and the overall outburst has shown two continuous flaring episodes with different profiles, spanning over two years. The X-ray flare was first detected by SRG/{\it eROSITA} during the second episode and had been continuously monitored by \sw and \Ni. It is unclear when the X-ray activity was triggered but several works have suggested that it was later than the optical (\citealt{Malyali2021,Chen2022,Wang2023}). The ultra-soft X-ray spectrum and optical spectral lines of {\avd} are consistent with a TDE, but the two consecutive optical flares are atypical of TDEs. In addition, \cite{Wang2023} reports the radio detection of this transient with VLA and VLBA, suggesting the possible ejection of a compact radio outflow (jet or wind) when the sources moved to a low-luminosity state.

The multi-wavelength study of \avd has been reported in \cite{Wang2023}. In this work, we focus on exploring in depth its X-ray temporal and spectral properties with proprietary {\Ni} and {\it Chandra} observations, and the archival data from {\sw}, which were performed $\sim459$\,days after the ZTF detection. 
The paper is structured as follows: in Section~\ref{sec:observation}, we describe our observations and data reduction; in Sections~\ref{sec:results} and \ref{sec:diss}, we present and discuss the overall evolution of the X-ray properties, respectively, and in Section~\ref{sec:conclu}, we conclude and highlight the main results of our work.

\section{Observations and data reduction}\label{sec:observation}
In this paper, uncertainties and upper/lower limits are quoted at the 1$\sigma$ and 3$\sigma$ confidence levels respectively. We adopt a redshift of 0.028 based on the report from the Transient Name Server\footnote{\url{https://www.wis-tns.org/object/2019avd}},
a luminosity distance of $D=130$\,Mpc from \cite{Wang2023} and a black-hole mass of $M_{\rm BH}$ of $10^{6.3}\,M_{\odot}$ from \cite{Malyali2021}.
We also adopt the Galactic absorption of $2.4\times 10^{20}\,\rm cm^{-2}$ from the HI4PI survey \citep{HI4PI} as the lower limit of the column density of {\avd}.

\subsection{XRT/{\sw}}
{\sw} has performed 51 observations on this source from May 13, 2020 to May 26, 2022. We used the 45 X-ray Telescope (XRT) observations which observed in photon counting mode, with total exposure time of 56.4\,ks. 

The XRT data were reduced with the tasks {\sc xrtpipeline} and {\sc xselect}. The source and background events were extracted using a circular region of 40\arcsec~and an annular ring with inner and outer radii of 60\arcsec~and 110\arcsec, respectively, both centered at the position of the source. The Ancillary Response Files (ARFs) were created with the task {\sc xrtmkarf} and the Response Matrix File (RMF) used was swxpc0to12s6\_20130101v014.rmf, taken from the Calibration Data base\footnote{\url{https://heasarc.gsfc.nasa.gov/docs/heasarc/caldb/swift}}.
Due to the small numbers of counts, the XRT spectra were grouped to have a minimum of 3 counts\footnote{\url{https://giacomov.github.io/Bias-in-profile-poisson-likelihood}} per bin using the FTOOL {\sc grppha}. Consequently, W-stats was used for the spectral fitting.
Since the source is background-dominated above 2\,keV, we only fitted the XRT spectra in the 0.3--2\,keV, which can be described well with an absorbed blackbody component until the late-time (MJD~59483) of the flare. Additionally, we calculate the hardness ratio of the count rates in the 0.8--2.0\,keV over that in the 0.3--0.8\,keV.

\begin{figure*} 
\centering  
\resizebox{2\columnwidth}{!}{\rotatebox{0}{\includegraphics[clip]{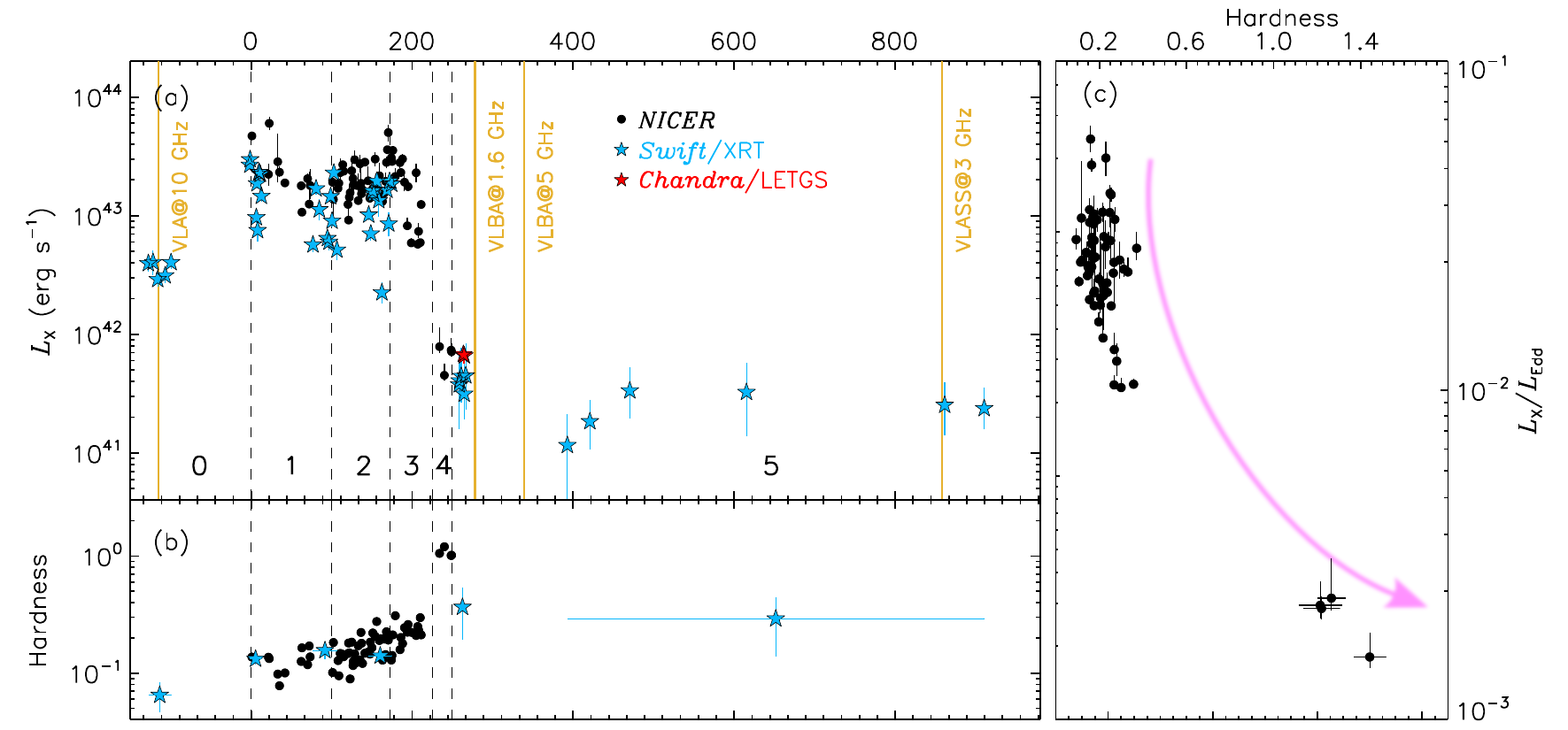}}}\vspace{-0.2cm}
\resizebox{2\columnwidth}{!}{\rotatebox{0}{\includegraphics[clip]{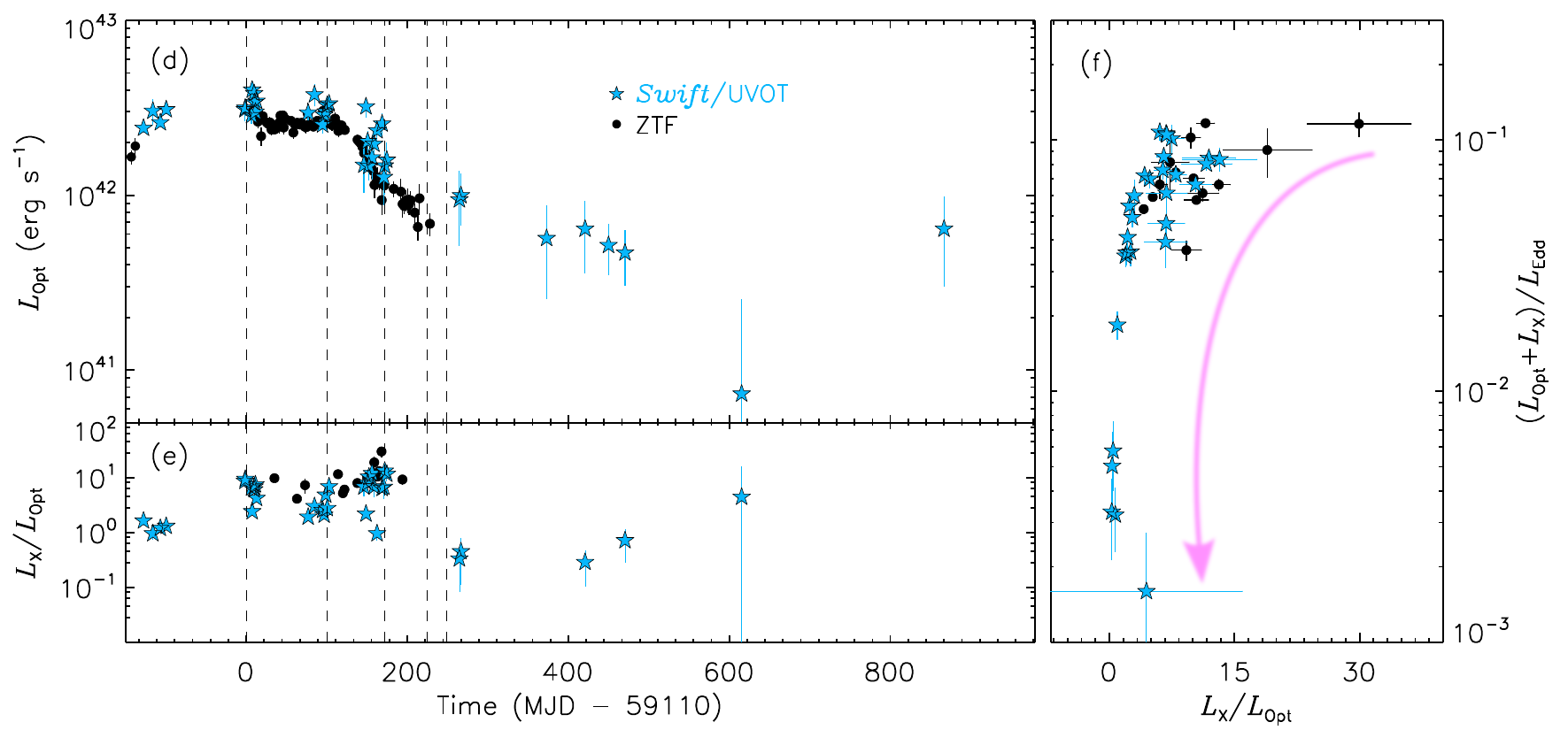}}}
\caption{\textbf{a:} Unabsorbed X-ray luminosity in the 0.3--2\,keV band. 
The black dots, the light blue stars and the red stars represent the {\Ni}, XRT/{\sw} and LETG/{\it Chandra} data, respectively. The yellow vertical lines indicate where there are radio detections (taken from \citealt{Wang2023}, except the last VLASS detection). The vertical dashed lines divide the flare into six phases, each identified by a number. 
\textbf{b:} The evolution of hardness ratio of the 0.8--2.0\,keV count rate with respect to the 0.3--0.8\,keV count rate (background excluded). To increase the signal-to-noise ratio of the XRT hardness ratio, we combined the data into six groups.  
\textbf{c:} Hardness-intensity diagram using {\Ni} data.
\textbf{d:} Optical and UV lightcurves. The black dots and blue stars the optical data in the {\it g} band of ZTF and the UV data in the UVW1 band of UVOT/{\sw}, respectively. The optical and UV data are taken from \protect\cite{Wang2023}, both of which are host-contribution subtracted and Galactic-extinction corrected.
\textbf{e:} The X-ray-to-optical/UV ratio. The dots and stars represent the ratios of the {\Ni} to the ZTF-{\it g} band and of the {\sw}-XRT to the UVOT-UVW1 band.
\textbf{f:} The X-ray-to-Optical/UV ratio vs the bolometric luminosity.
The light blue arrows denote the upper limit of the detection at a 3$\sigma$ confidence level. The magenta arrows indicate the temporal evolution of the ratios.}
\label{fig:lc}
\end{figure*}

\subsection{LETG/{\it Chandra}}
{\avd} was observed twice with {\it Chandra} Low Energy Transmission Gratings (LETGs) on June 8 and 9, 2021 (ObsID 25056 and 25060; PI: Pasham) with total exposure time of 50\,ks. The data were reduced using {\sc ciao}\footnote{\url{https://cxc.cfa.harvard.edu/ciao}} version 4.13 and CALDB version 4.9.5. We first reprocessed the data with the script {\sc {\it Chandra}\_repro} to generate new level-2 event files. 
We then ran the tool {\sc tgdetect} to determine the source position in the {\it Chandra} image. However, we detected no source with significance $> 2\sigma$, which suggests that the source is too faint for a further spectral analysis. Alternatively, we run {\sc srcflux} to estimate the source flux centered on the source location by assuming an absorbed \texttt{powerlaw}\footnote{As reported by \cite{Wang2023} and our analysis on the \sw and \Ni data, the spectrum hardened as the luminosity decreased. We hence assumed a powerlaw spectrum to estimate the current flux.} spectrum with $\Gamma =2$ in the 0.7--10\,keV band. The value of $\Gamma$ is inferred from the latest {\Ni} spectrum. We obtained a luminosity of (5.7--8.0)$\times10^{41}$\,\ergs at the time of the {\it Chandra} observations.

\subsection{XTI/\Ni}\label{sec:nicer_datareduction}
{\Ni} performed high-cadence monitoring observations of {\avd} with the X-ray Timing Instrument (XTI) from September 19, 2020 until June 16, 2021, split across 207 observations with a cumulative exposure time of approximately 408.5\,ks. 
We reprocessed the data using {\sc nicerdas} version 2020-04-23 and CALDB version xti20200722. In addition to the standard data reduction steps to filter, calibrate and merge the {\Ni} events, we excluded some XTI detectors on an observation basis when the count rates deviate $> 3\sigma$ from the mean, or which are switched off during observation. Besides that, FPMs~14 and 34 are always excluded since they often exhibited episodes of increased detector noise. 
We applied the tool {\sc nibackgen3C50} \citep{Remillard2021} with level 3 filtering criteria, i.e. $hbgcut=0.05$ and $s0cut=2$, to extract and estimate {\Ni} spectra and background. The RMFs and ARFs tailored for the selected detectors were applied in the spectral analysis. Finally, all the spectra were grouped with {\sc ftgrouppha} using the optimal binning scheme \citep{Kaastra2016}. 
We also excluded some {\Ni} observations when i) the exposure time was shorter than 100\,s; ii) the background and/or the contamination (see the next paragraph) was higher than the source flux above 1.7\,keV; iii) the observations were performed after MJD~59360 (these observations are mostly contaminated by Sun glare). In the end we obtained 153 {\Ni} observations left for further analysis. Similar to the XRT spectra, there is no significant emission above 2\,keV so we only fit the {\Ni} spectra in the 0.3--2\,keV band. We also calculated the hardness ratio with the {\Ni} data using the same energy bands applied to the XRT data.

Additionally, there is another persistent X-ray source, IC~505, at a distance of about 3.7\arcmin~from {\avd} in the large Field of View (FoV) of XRT. 
We compare the position of this contaminating source with {\Ni}'s FoV with a radius of 3.1\arcmin~\citep{Wolff2021,Pasham2021} and find that this source is located on the edge of the {\Ni}'s FoV. As observed by both \xmm/EPIC-pn in 2015 and \sw/XRT in 2019--2022, IC~505 appears to be stable over time (see Fig.~\ref{figA:app_image}), with a \texttt{powerlaw} spectrum plus an emission line around 1\,keV, whose flux is $\sim 0.3\times10^{-12}\rm \,erg\,cm^{-2}\,s^{-1}$ in the 0.3--2\,keV. Based on this flux level, we determined that part of the emission in \Ni observations of \avd is contaminated by IC~505. However, owing to the stability of the flux of IC~505, we were able to subtract its contribution from the \Ni data of \avd (the method is described in detail in the Appendix~\ref{sec:app_contamination}).


\section{Results}\label{sec:results}
\subsection{Flare evolution}
We show the long-term X-ray lightcurve of \avd in Fig.~\ref{fig:lc}a. 
The {\Ni}, {\sw} and {\it Chandra} data are labeled as black dots, light blue and the red stars, respectively. The first detection of the {\Ni} campaign, MJD~59110, which refers to the peak of the flare in X-rays, is defined as day 0. 
In Fig.~\ref{fig:lc}b, the black dots and the light blue stars denote the hardness ratio derived from the {\Ni} and {\sw} observations, respectively.

To better describe and locate the evolution of the flare in different periods, we divided the entire lightcurve into six phases which are marked with Arabic numerals from 0 to 5 and are separated with vertical dashed lines in Fig.~\ref{fig:lc}.
Phases 0 and 5 correspond to the periods prior to and after the flare, respectively without \Ni observations, while phases 1 till 4 correspond to Days 0--100, 101--172, 173--225 and 226--249, respectively. The evolution of the flare can be described as follows.

In phase 0, the luminosity first increased by more than an order of magnitude and peaked around Day 0. In phase 1, the luminosity decreased by nearly a factor of 5 while the hardness ratio remained constant. During phase 2, the luminosity reached another peak and the hardness ratio gradually increased. In phase 3, the luminosity decreased rapidly by an order of magnitude and the hardness ratio increased. Later in phase 4, the source continuously dimmed while the hardness ratio increased. 
Right after phase 4, {\avd} went into Sun glare afterwards for {\Ni}. 
Follow-up XRT and LETG observations between Days 258 and 267, 9\,days after phase 4, show a further decrease in the source flux. After the seasonal gap (between Days 266 and 373) in phase 5, the X-ray luminosity remained the same as in phase 4 and lasted for over $\sim$700\,days, the flux being more than two orders of magnitude lower than the peak of the flare in phase 1.

Compared to \Ni, XRT has a relatively smaller effect area; also, the XRT observations were either of relatively short exposure time or were performed when the source was relatively faint ($L_{\rm X}<10^{43}$\,\ergs). 
To increase the signal-to-noise of the hardness ratio, we divided the whole XRT dataset into six groups based on the observation time and combined the data from each group to calculate the hardness ratio. We added the XRT hardness as light blue stars to Fig.~\ref{fig:lc}b, in which the error bars along the x-axis indicate the duration of each group. The XRT hardness shows a comparable evolution with the {\Ni} hardness and further increases after the {\Ni} campaign (we only compare the hardness ratio derived from the same instrument).

The {\Ni} hardness intensity diagram (HID) is shown in Fig.~\ref{fig:lc}c: the hardness ratio remained nearly constant as the luminosity decreased in phases 1--3 and then increased as the luminosity further decreased in phase 4. We convert the luminosity to units of the Eddington luminosity ($L_{\rm Edd}$), which peaks around 0.24\,$L_{\rm Edd}$.

Fig.~\ref{fig:lc}d shows the optical and UV photometries derived from the $g$ band of ZTF and the UVW1 band of {\sw} with symbols of dots and stars. These data are taken from \cite{Wang2023}, which have been corrected for the Galactic extinction and contribution from the host galaxy. 
Fig.~\ref{fig:lc}e shows the X-ray to optical/UV luminosity ratio: the dots represent the ratio of luminosities inferred from {\Ni} and ZTF, and the stars represent the ratio of luminosities inferred from XRT and UVOT. The trends of the two curves are consistent when the observations are quasi-simultaneous. Fig.~\ref{fig:lc}f shows the X-ray-to-optical ratio as a function of the bolometric luminosity which is taken as the sum of the optical/UV and the X-ray luminosities. Similar to the trend of the HID curve shown in Fig.~\ref{fig:lc}c, the X-ray-to-optical ratio decreases as the bolometric luminosity decreases.

\begin{figure}
\centering  
\resizebox{0.9\columnwidth}{!}{\rotatebox{0}{\includegraphics[clip]{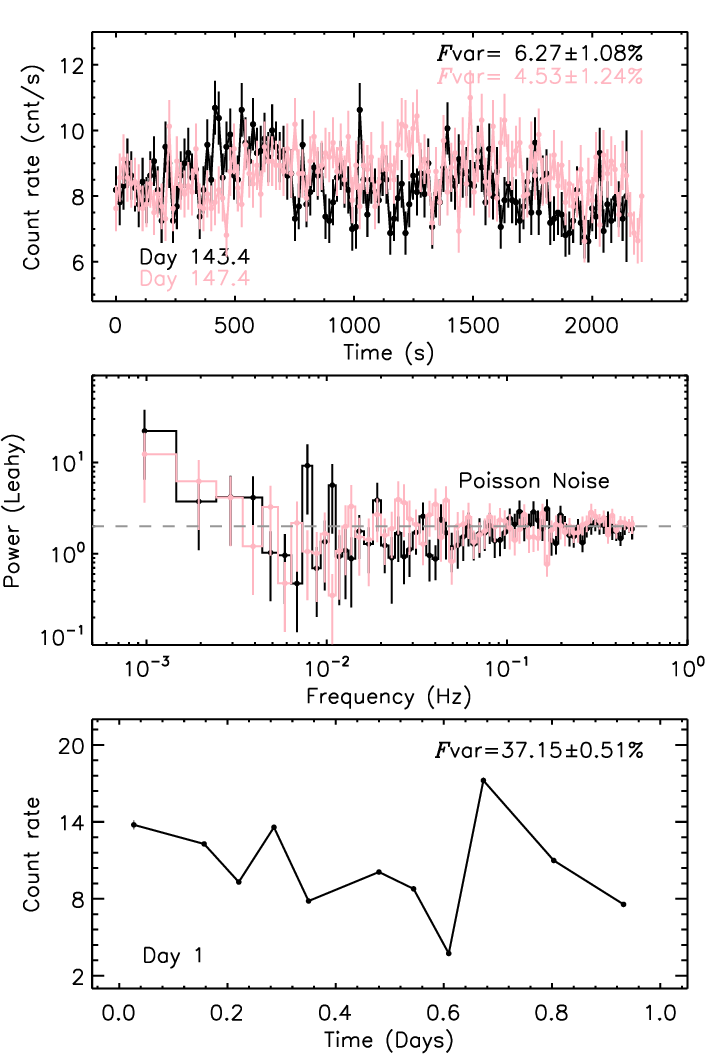}}}
\vspace{-0.3cm}
\caption{\textbf{Top:} Examples of lightcurves in bins of 16\,s. \textbf{Middle:} The corresponding periodogram of the above lightcurves. The dashed line shows a Poisson noise level of 2. \textbf{Bottom:} An example of lightcurve in bins of GTI.}
\label{fig:lc_psd}
\end{figure}

\subsection{X-ray timing properties}\label{sec:result_vari}
As shown in Fig.\ref{fig:lc}a, the X-ray emission is highly variable.
To quantify the variability, we first generated power spectral densities (PSDs) with Leahy normalisation \citep{Leahy1983}. Since the duration of the \Ni Good Time Intervals (GTIs) ranges between 18 and 2,146\,s, the frequency of the periodogram can only be traced down to $\sim0.5$\,mHz. We show two examples of the lightcurves with longest duration and the corresponding periodogram with a frequency range of $1-500$\,mHz in Fig.~\ref{fig:lc_psd}. There is no periodic signal present in either periodogram, but the power of the periodogram tend to increase towards lower frequencies. As a comparison, we show an example of the lightcurve spanning roughly one day on the bottom panel of Fig.~\ref{fig:lc_psd}, in which the flux variation is much larger. These suggest that the variability of \avd dominates over longer timescales.

To further explore the short-term variability in a longer duration, we generated the background-subtracted lightcurve in bins of GTI and then computed the excess fractional rms variability amplitude, $F_{\rm var}$, between GTIs of each observation (see Appendix~\ref{sec:app_rms} for more details on the calculation). 
To reduce the bias of the length of individual observation, we chose only observations lasting longer than 10\,ks with more than 4 GTIs. 
Equation~\ref{equ:excess} shows that $F_{\rm var}$ of each individual observation can only be computed if the variance is larger than the measured errors. Thus we ignored the data that do not fulfill this criteria. 
The absolute rms amplitude, $\sigma_{\rm XS}$, was computed as $F_{\rm var}$ multiplied by the mean count rate of an observation.

\begin{figure*}
\centering  
\resizebox{0.78\columnwidth}{!}{\rotatebox{0}{\includegraphics[clip]{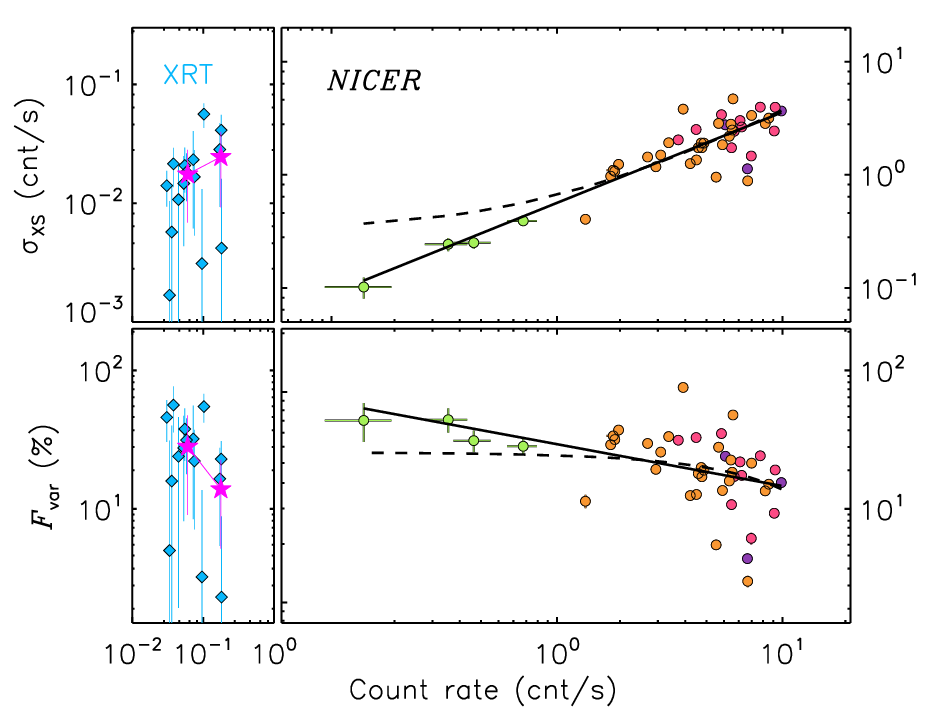}}}
\hspace{-0.2cm}
\resizebox{0.68\columnwidth}{!}{\rotatebox{0}{\includegraphics[clip]{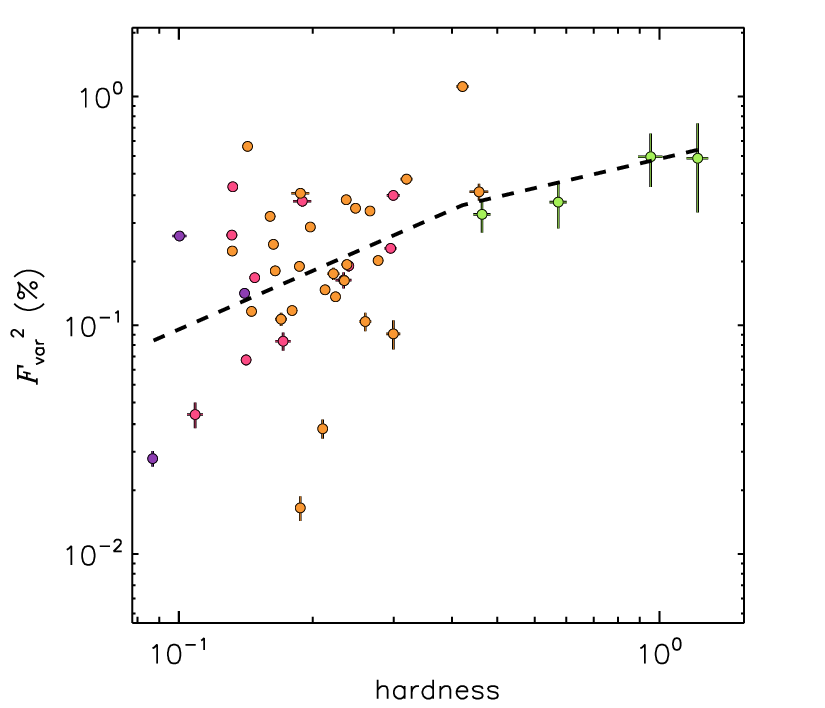}}}
\hspace{-0.4cm}
\resizebox{0.68\columnwidth}{!}{\rotatebox{0}{\includegraphics[clip]{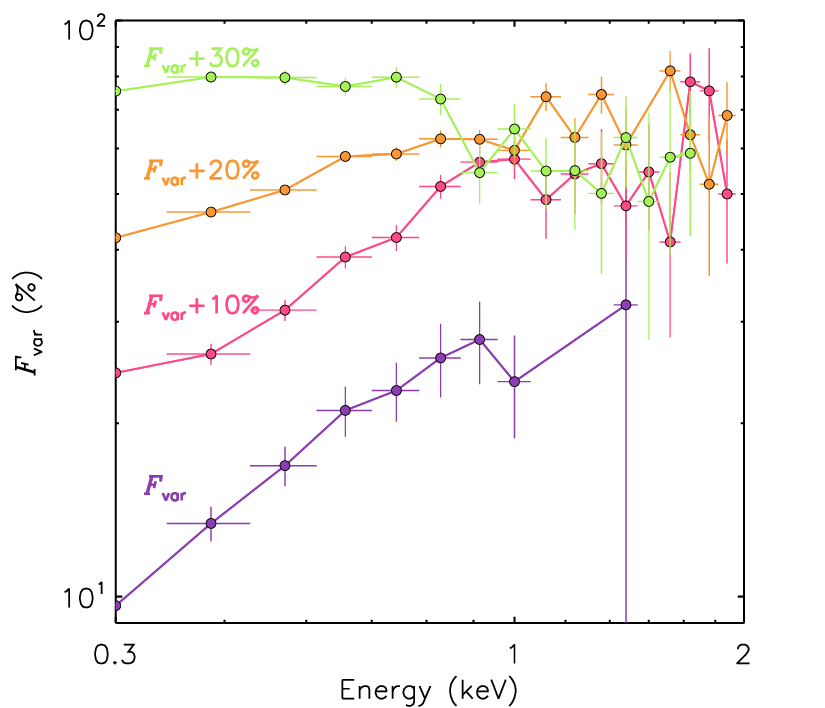}}}
\vspace{-0.1cm}
\caption{\textbf{Left panel:} Absolute and fractional rms amplitude vs count rate. The magenta stars represent the rebin of the XRT data. The solid and dashed lines indicate the power-law and linear fits to the data of \Ni, respectively. The purple, magenta, yellow and green dots correspond to phases 1--4, respectively.
\textbf{Middle panel:} fractional rms amplitude vs hardness.
\textbf{Right panel:} fractional rms amplitude vs energy. The offset of $F_{\rm var}$ is labeled legend.}
\label{fig:rms}
\end{figure*}

As illustrated in Fig.~\ref{fig:lc_psd}, the variability in the time scale of minutes is significantly higher than in that of seconds, e.g. $F_{\rm var}$ increasing by a factor of 6--8. We further plot $\sigma_{\rm XS}$ and $F_{\rm var}$ as a function of the count rate in Fig.~\ref{fig:rms}.
$F_{\rm var}$ is quite high and scattered, ranging between 12.6\%--105.1\% with an average of 47\% (42\% in phases 1--3 and 65\% in phase 4). 
$\sigma_{\rm XS}$ is strongly correlated to the count rate while the correlation between $F_{\rm var}$ and count rates shows an opposite trend, $F_{\rm var}$ decreasing with increasing flux.
We fitted a power-law and a linear model to each curve of \Ni shown in Fig.~\ref{fig:rms} and found that the dots in phase 4 deviate from the linear fit of both curves.

To test the validity of the high $F_{\rm var}$, we applied the same method to the XRT data. Given that most of the XRT observations consist only of one GTI, we computed $F_{\rm var}$ of the XRT data within a GTI (in bins of 128\,s). Due to the low statistics, we were only able to measure $F_{\rm var}$ of 16 GTIs in phases 1--3. We added the results as light blue diamonds together with measurements of the average (marked as magenta stars in Fig.~\ref{fig:rms}). Although with considerable uncertainties, $F_{\rm var}$ values derived from the XRT data are consistent with the \Ni data. 
Even though the $F_{\rm var}$ values from the \Ni data could be over/underestimated as the GTIs are not evenly sampled, the comparable values of $F_{\rm var}$ provided by XRT support the high variability detected by \Ni.

To show the relation between $F_{\rm var}$ and the spectral state, we plot $F_{\rm var}^2$ against spectral hardness in the middle panel of Fig.~\ref{fig:rms}, in which $F_{\rm var}^2$ increases with the hardness. 
The overall curve can be described by a broken power-law with two regions and the inflection point corresponds to where the state transition occurs. More specifically, the left side (hardness $<$ 0.4) of the curve is relatively soft and occupied by the data in phases 1--3, with an increase in variability with increasing hardness; while the right side (hardness $>$ 0.4) is hard and occupied by the data in phase 4, with a further increase in variability as the spectrum hardens. 

To further explore the energy dependence of the variability, we computed $F_{\rm var}$ spectra and illustrated it in the right panel of Fig.~\ref{fig:rms}. 
In phases 1--3, $F_{\rm var}$ increases with energy until 1\,keV and then show some wiggles at higher energies, indicating that the rms spectrum consists of two components. Conversely, the value of $F_{\rm var}$ in phase 4 is higher than at the other phases and show invariant evolution at energies below 1\,keV and it diminishes at higher energies.

\subsection{X-ray spectral properties}
We study the flux-averaged X-ray spectrum in units of observation in this section.
The spectrum is ultra soft, being background dominated above 2\,keV during the flaring episode. The XRT spectrum can be reasonably described by an absorbed \texttt{bbody} until phase 5, when an absorbed \texttt{powerlaw} is required instead. Compared to XRT, the {\Ni} spectrum shows more complex features, requiring two components in most cases. As the best-fitting parameters of the XRT spectrum have already been shown in \cite{Wang2023}, we focus on the {\Ni} spectra in this work.

To describe the \Ni spectra, we employed a phenomenological model including one absorbed \texttt{bbody} component plus a \texttt{powerlaw} component. The redshift of the galaxy was taking into account by an additional \texttt{zashift} component which has been fixed at 0.028.
In addition to the two continuum components, there is an emission-line feature below 1\,keV or absorption-line feature above 1\,keV in individual observations. 
Unfortunately, the emission line is also present in the spectrum of IC~505, making it difficult to associate it \avd and hence we do not investigate it further.
As for the absorption line, we show an example of the unfolded spectrum in Fig.~\ref{fig:abs_compare} without and with the contamination from IC~505. An absorption line with a centroid energy of 1.03$\pm$0.01\,keV is clearly present in the spectrum, which may indicate the presence of disk winds.

Fig.~\ref{fig:abs_compare} also shows the continuum components. The soft excess, modeled with \texttt{bbody}, is significantly required in phases 1$-$3. In phase 4, only one component is required. Either a \texttt{powerlaw} or a \texttt{bbody} can describe the data but the former always provides a statistically better fit than the latter, i.e. obtaining a smaller $\chi^2$.
This has also been mentioned in \cite{Wang2023}, i.e. they were unable to statistically distinguish whether the spectrum was thermal or non-thermal at the beginning of phase 5; however, the spectrum flattened clearly later (i.e. Days 373-615, see Fig.~4 of \citealt{Wang2023}).
These evidence suggests that the source evolved from a \texttt{bbody}-dominated/soft state (phases 1--3) to a \texttt{powerlaw}-dominated/hard state (phases 4-5). The state transition occurred around $L_{X}\sim0.01\,L_{\rm Edd}$.

\begin{figure} 
\centering  
\resizebox{0.9\columnwidth}{!}{\rotatebox{0}{\includegraphics[clip]{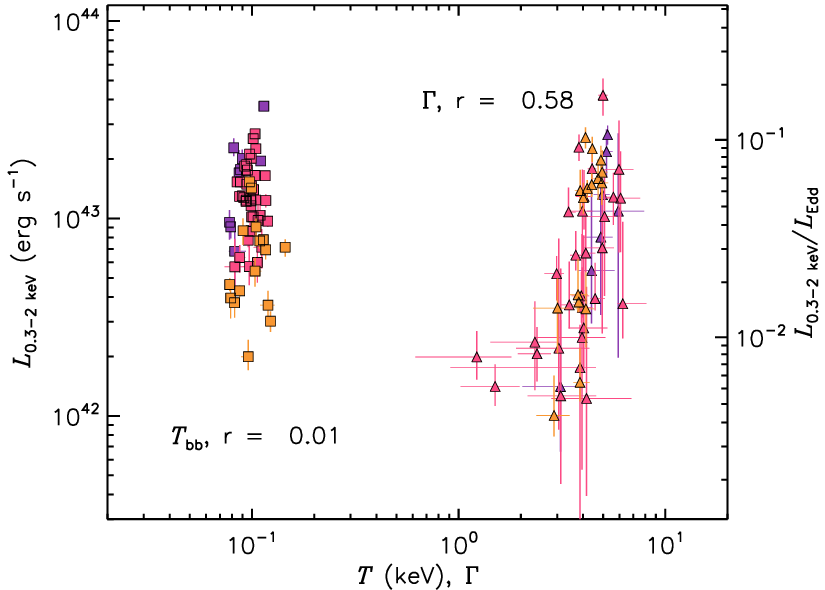}}}
\caption{Blackbody temperature/photon index versus the \texttt{bbody} and the \texttt{powerlaw} luminosity, respectively. $r$ represents linear Pearson correlation coefficient. The colors are defined the same as in Fig.~\ref{fig:rms}.}
\label{fig:lumi_spec}
\end{figure}

We plot the evolution of the best-fitting parameters and the individual flux of the continuum components in Fig.~\ref{fig:para_pl}. The full description of the spectral evolution is present in App.~\ref{sec:app_spec}.
We further plot the blackbody temperature and the photon index in phases 1--3 as a function of the luminosity in Fig.~\ref{fig:lumi_spec}. As the luminosity decreases by over one magnitude, the blackbody temperature remains constant and the photon index decreases with the luminosity.

Additionally, to explore the properties of any outflowing materials via the absorption feature, we generated a grid of photoionisation models with varying column density ($N_{\rm gas}$) and ionisation parameter ($\log\,\xi_{\rm gas}$) with \textsc{xstar} \citep{Kallman2001}.
We assumed a \texttt{bbody} spectrum with a temperature of 0.1\,keV and a luminosity of $2.0\times 10^{43}\rm \,erg\,s^{-1}$ (integrated from 0.0136 to 13.6\,keV), irradiating some materials with a fixed gas density of $10^{10}\rm \,cm^{-3}$, a turbulent velocity of $10^5$\,\kms (based on the width of the line) and solar abundances. We replaced \texttt{gaussian} with \texttt{xstar} and obtained a fit with $\rm \chi^2/\nu=20.25/19$. The latter reveals $N_{\rm gas}=2.5\pm0.2\times 10^{23}\rm \,cm^{-2}$, $\log~\xi_{\rm gas}=3.44_{-0.12}^{+0.28} \rm\,erg\,cm\,s^{-1}$ and a blueshifted velocity of $0.08\pm0.02$c. Hence the absorption feature could be due to Ne~{\sc ix}, Fe~{\sc xix}, Fe~{\sc xx} or a mixture of several of them. Further discrimination cannot be achieved with the spectral resolution of the data.

\section{Discussion} \label{sec:diss}
We have used the {\Ni}, {\sw} and \textit{Chandra} observations to study the accretion and the ejection properties of the highly variable nuclear transient {\avd}. The X-ray flaring episode has lasted for over 1000\,days, spanning over two orders of magnitude in luminosity. 
A rapid decrease in the luminosity has been observed at $\sim$225\,days after the peak of the X-ray flare, followed by a state transition occurring at 0.01\,$L_{\rm Edd}$.
In the following, we discuss the X-ray variability, the possible physical origins of the soft excess with a constant temperature, the potential triggers for the state transition, and eventually compare our target with other accreting black holes.

\subsection{Evolution of the X-ray variability}\label{sec:vari}
Linear absolute rms-flux relation has been observed in all types of accreting systems \citep[e.g.][]{Vaughan2003,Uttley2001,Heil2010,Scaringi2012} and has been interpreted as a result of inwardly propagating variations via accretion flows \citep{Lyubarskii1997}. In the case of \avd, although with some scatters the absolute rms-flux relation shows deviations from linearity in phase 4 (see Fig.~\ref{fig:rms}). Together with the fractional rms-flux relation, it suggests that either the intrinsic variations of the emission in phases 1--3 and 4 are different and/or the geometry of the accretion flow has changed, e.g. from a slim to a thick disk. We discuss these possibilities below.

As shown in Fig.~\ref{fig:rms}, the fractional rms amplitude $F_{\rm var}$ of \avd is very high, with an average of 43\% and its evolution is related to the spectral state. Such a high variability has only been observed in limited accreting systems, for instance BHX GRS~1915+105 \citep{Fender2004} and ULXs NGC\,5408~X--1 and NGC\,6946~X--1 \citep{Middleton2015,Atapin2019}. The common properties of these targets are super-Eddington accreting rates with strong outflows. 
Although the X-ray luminosity of \avd is below the Eddington luminosity, based on the fit to the broadband spectral energy distribution (SED) from optical to X-rays, \cite{Wang2023} found the bolometric luminosity to be 6.7\,$L_{\rm Edd}$\footnote{It is worth noting that the absence of extreme UV emission results in a consistent underestimation of observed luminosity in both the optical/UV and X-ray spectra. However, it is crucial to note that the SED modeling discussed in \cite{Wang2023} incorporates significant absorption at the \avd location, a factor that is not considered in the current study. Therefore, caution is advised when interpreting the absolute value of the bolometric luminosity in this context.}, suggesting that \avd is in the super-Eddington regime in phases 1--3. 


\cite{Middleton2015} has proposed a spectral-timing model taking into account both intrinsic variability via propagated fluctuations and extrinsic variability via obscuration and scattering by winds to explain supercritically accreting ULXs.
In this scenario, the evolution of the overall variability and the spectral hardness are suggested to be determined by the changes in two parameters: mass accretion rate and inclination angle. The latter may vary due to precession of the accretion disk.
Under this framework, the evolution of $F_{\rm var}$ against hardness of \avd is consistent with the mass accretion rate deceasing while the inclination angle either decreases or remains constant. 
More explicitly, when the disk inclination angle is moderate\footnote{If the scale height of the disk is large, a small inclination angle may also work.} and the mass accretion rate is high, disk winds would have a small opening angle towards the observer and thus have a higher probability to intercept the high-energy emission from the hot inner disk. This would result in a softer spectrum and highly variable hard X-rays via obscuration and scattering. Whilst either the mass accretion rate and/or the inclination angle decrease, the opening angle of the disk winds towards the observer increases and more of the hot inner disk would expose directly to the observer. Subsequently, the dominant emission will harden and exhibit reduced variability due to the decrease in obscuration and scattering, at odds with the further increase of the variability and the dimming of the target in phase 4. Even in phase 5, \avd remains highly variable (see the green dots in the rightmost panel of Fig.~\ref{figA:app_image}). 
In fact, ULXs are persistent sources such that we should not expect transients like \avd to share all the properties of ULXs, especially after the abrupt decrease in the mass accretion rate in phase 4.

To examine whether the variability can be attributed to the existence of a local absorber, potentially leading to increased absorption or obscuration, we divided the XRT data in phases 1--3 into two segments based on the luminosities exceeding and falling below $10^{43}\,\rm erg\,s^{-1}$.
Then we jointly fitted the spectra from the two segments with an absorbed \texttt{bbody} component plus a \texttt{powerlaw} component. To improve the constraint on $N_{\rm H}$, both $kT_{\rm in}$ and the \texttt{powerlaw} component are linked to vary across observations. 
We verified the necessity of the power-law component for both segments using the {\sc ftest} command. The null probability for the high-flux segment is $9.4\times10^{-9}$, and for the low-flux segment, it is $7.7\times10^{-5}$.
Next, we compared the values of $N_{\rm H}$ obtained from the fit with and without the \texttt{powerlaw} component. Interestingly, we found that for the low-flux segment, $N_{\rm H}$ tends to be slightly higher, while the powerlaw flux is lower. However, the disparity in $N_{\rm H}$ values between the two segments remains consistent within uncertainties, specifically $1.03\pm0.52\times10^{21}\,\rm cm^{-2}$ and $0.63\pm0.28\times10^{21}\,\rm cm^{-2}$.
In summary, these results may suggest the presence of a local absorber, but a more definitive conclusion is hampered by the data quality.

Alternatively, the evolution of $F_{\rm var}$ against hardness/count rate is consistent with the one seen in BHXRBs when a source evolves from a relatively soft to a hard state during the decaying phase of an outburst (e.g. \citealt{Stiele2017,Wang2020,Alabarta2022}).
The temporal evolution of the hardness and the flux of \avd, as well as the HID curve and the radio emission detected in phase 5, are also in a good agreement with the decaying phase of an outburst and the launch of a jet of BHXRBs. Considering the data used for the computation of $F_{\rm var}$, the corresponding frequency range would be roughly 10$-$100\,$\mu$Hz, which could be converted to 1$-$10\,Hz for a stellar-mass BH of $10\,M_{\odot}$. However, the fraction rms in such a frequency range of BHXRBs is normally 15--30\% (e.g. \citealt{Heil2012,Wang2020}). 
This unusually high rms makes \avd different from typical sub-Eddington accreting BHs.

\begin{figure} 
\centering  
\resizebox{0.9\columnwidth}{!}{\rotatebox{0}{\includegraphics[clip]{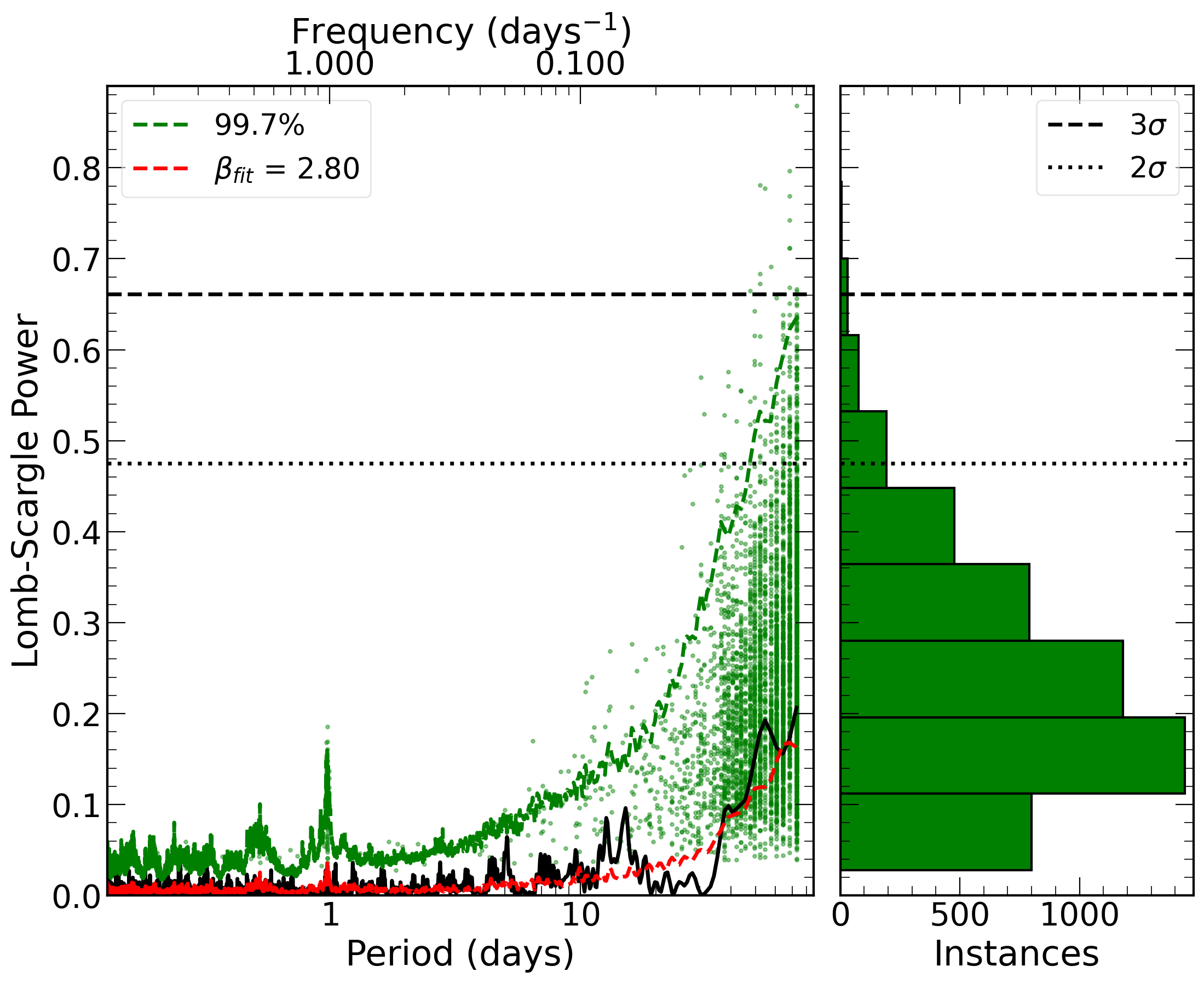}}}
\caption{Lomb-Scargle periodogram of the \Ni data (solid black line) with the green dashed line showing the one-trial 3$\sigma$ false alarm probability, while the multiple trials 2$\sigma$ and 3$\sigma$ false-alarm probability is shown in black dotted and dashed lines respectively. The red line represents the average of the simulated periodograms. The scatter green dots and the right panel show the position (in frequency) and the distribution of maximum peaks in the simulated lightcurves.}
\label{fig:lomb-scargle}
\end{figure}

Short-term variability on timescales of hundreds to thousands of seconds has been explored in several TDEs, such as Swift~J1644+57 \citep{Saxton2012b,Reis2012}, ASASSN--14li \citep{Pasham2019}, etc.
Such short timescales indicate that the X-ray emitting region of TDEs is compact, e.g. $3\times10^{12-13}$\,cm.
\avd tends to show both small variations on timescales of hundreds of seconds (see Figs.~\ref{fig:lc_psd}) and large variations on timescales of thousands of seconds (see \ref{fig:rms}). 
The latter is revealed by the presence of dips.
\cite{Saxton2012b} observed a dipping behaviour with peculiar patterns in the brightest relativistic TDE Swift~J1644+57. The spectrum softened during the dips without requiring changes in column density. This evidence support their idea that the dips were driven by the precession and nutation of jets.
 
Recently, \cite{Chen2022} proposed that the X-ray variability of \avd could be due to Lense–Thirring precession of the accretion disc, and predicted a precession period of 10--25\,days. In order to test whether the variability is associated with precession, we conducted lightcurve simulations to infer the power spectral densities of the source (taking into account gaps, aliasing and red-noise leakage effects) and used such estimate (and its uncertainties) as the null-hypothesis to test for the presence of any significant peak in the (Lomb-Scargle) periodogram. 
We refer the reader to Appendix~\ref{sec:app_ls} for the details. The results are shown in Fig.~\ref{fig:lomb-scargle} which show the 3$\sigma$ one- (blue dotted line) and multiple-(green dashed line) trial(s) false alarm probabilities based on the inferred null-hypothesis (red line). Overall, we find no obvious peak above the false alarm probability levels. This is supported by the good agreement between the best-fit periodogram (red line) and the data (with a rejection probability of ~50\%; see Appendix~\ref{sec:app_ls}). We thus conclude that the variability of \avd is fully consistent with the typical aperiodic variability commonly observed in accreting systems. 

Overall, we conclude that the high variability of the X-rays of \avd are very likely due to the presence of clumpy winds, and the state/luminosity dependent evolution of the variability in phases 1--3 could be interpreted by the decrease in the mass accretion rate and subsequent opening angle of the funnel of the supercritical disk. It remains unclear why \avd remains so variable even after the luminosity drops below 1\%\,$L_{\rm Edd}$. It however should be noted whether supercritical accretion can initiate clumpy winds is a topic of ongoing debate. Some 2D numerical simulations (e.g. \citealt{Takeuchi2013,Takeuchi2014}) propose that super-Eddington clumpy winds might be induced by the Rayleigh-Taylor instability. On the other hand, some 3D simulations employing radiation magnetohydrodynamics or relativistic radiation magnetohydrodynamics (e.g. \citealt{Jiang2014,Sadowski2016}) suggest the opposite.
The disk precession scenario cannot be completely ruled out but its contribution to the observed variability should be limited.

\subsection{Constant blackbody temperature}
The evolution of thermal blackbody temperatures in TDEs has been extensively studied in the optical-UV band, and they have shown different trends along with the flare (e.g. \citealt{Hinkle2020,van_Velzen_21,Hammerstein2022}). Some of the temperatures are observed to remain roughly constant with small-scale variations, e.g. ASASSN--14li \citep{Holoien2016a} and ASASSN--18pg \citep{Holoien2020}, and some show either a decrease or increase trend \citep{van_Velzen_21}.
In one of the best observed TDEs ASASSN--14li, while the luminosity dropped nearly two orders of magnitude over 600\,days, the blackbody temperature remained relatively constant in the optical/UV \citep{Holoien2016a} and decreased at most by a factor of 2 in the X-rays \citep{Brown2017}. 
Due to the scarce data sample of TDEs with X-ray emission, the evolution of the characteristic temperature in X-rays has not been statistically studied as much as in optical-UV bands.

A soft excess has been detected at least in phases 1--3 of the X-ray flare of {\avd}. 
Irrespective of its physical origin, the soft excess can be well described by a \texttt{bbody} component.
The obtained $T_{\rm bb}$ is rather stable during phases 1--3 while the luminosity decreases over one order of magnitude. 
Fig.~\ref{fig:lumi_spec} shows that the $T_{\rm bb}$-$L_{\rm bb}$ relation does not follow the $L\propto T_{\rm bb}^{4}$ scaling for a optically thick, geometrically thin accretion disk \citep{Shakura1973}. 
This means that the accretion disk of \avd is very unlikely to be thin.

A similar soft excess below 2\,keV has been ubiquitously observed in AGN with low column densities (e.g. \citealt{Singh1985}), although the origin of this excess remains debated. We compare our result with the two most popular models for the soft excess in AGN: warm, optically thick Comptonization (e.g. \citealt{Gierlinski2004,Petrucci2018}) and blurred ionized reflection (e.g. \citealt{Ross1993,Ballantyne2001,Kara2016}). 
To test the Comptonization scenario, we replaced the \texttt{bbody} component with a Comptonization component, \texttt{comptt}, and obtained a comparable fit. However, even for the spectra observed around the peak of the flare and with a long exposure time, the parameters of \texttt{comptt} cannot be constrained well.   
We then fixed $\Gamma$ of \texttt{powerlaw} to the value derived from the original model to avoid model degeneracy, e.g. in the fit to the spectrum of ObsID. 3201770101 (the first observation of the \Ni campaign with exposure time of 7.1\,ks), the best-fitting value of $\tau$ is up to 0.3 at $1\sigma$ confidence level, with the temperature of the seed photon and the hot plasma of $100-$102\,eV and $7.1-$10.8\,keV, respectively.
This value is much lower than the typical optical depth in AGN where $\tau=10-20$, indicating that the required Comptonization region is optically thin instead of thick and hence inconsistent with this scenario. 
In the blurred ionized reflection scenario, if the soft excess is produced by the disk reflection from hard X-rays, i.e. the \texttt{powerlaw} component in our case, the \texttt{powerlaw} and \texttt{bbody} fluxes should be correlated. Even though they appear to be correlated in phase 3, they seem to be marginally anti-correlated in phase 2 (see Fig.~\ref{fig:para_pl}), at odds with this scenario.
Overall, the soft excess observed in {\avd} cannot be solely interpreted by either of these two scenarios. 

In fact, since the inner hot X-rays could have been obscured and scattered into softer X-rays or even UV emission, the observed soft excess would actually greatly deviate from physical reality. Recently, \cite{Mummery2021} describes the impact of the use of single-temperature blackbody, including the disk inclination angle and the local absorber, such as stellar debris and outflows, on the determination of the disk radius and temperature in the context of TDEs. They suggest that the disk temperature would have been overestimated/underestimated if neglecting the effect of the hardening factors/the local absorber.
Due to the limited bandpass and the complexity of the spectrum of \avd, we do not explore more sophisticated models here and the present disk temperature should be taken with caution.
More data and theoretical work are needed to understand whether such a constant temperature in X-rays is due to observational/model effects and/or is driven by some physical mechanism, which are beyond the scope of this work.

\subsection{Rapid dimming in luminosity}
A fraction of TDEs experience abrupt dimming in X-rays, e.g. partial TDEs \citep{Wevers2021,Wevers2023,Liu2023} and jetted TDEs \citep{Zauderer2013}. Both of them present a similar rapid drop in X-ray luminosities as in \avd along with a state transition. 

Regarding partial TDEs, if one considers $\Gamma$ obtained from the \texttt{powerlaw} model, both the evolution and the values in \avd are comparable to those observed in the partial TDE eRASSt~J045650.3--203750 \citep{Liu2023}. However, after monitoring \avd for over 1,500\,days in optical, there is no sign of a second re-brightening (i.e. a third flare) as seen in partial TDEs. 
In fact, the optical evolution of \avd is more similar to that of a TDE with a successive re-brightening events (see examples in Fig.~9 of \citealt{Yao2023}), although \avd is the only TDE candidate for which the re-brightening is stronger than the initial one. In conclusion, the properties of \avd do not resemble those seen in on-axis jetted TDE or partial TDEs.

In \avd, the rapid dimming is followed by a soft-to-hard state transition and the spectrum continuously hardens. At the same time, 
the radio emission in phase 5 seems to increase. 
Similar phenomena have been observed in the micro-quasar GRS~1915+105. \cite{Motta2021} reports a strong radio flare accompanied by a significant reduction in X-ray activity and a hardening of the spectrum. In addition, a local absorber is required to describe the spectrum shape, so they attributed the reduction in X-ray flux to the high inhomogeneous absorption. Different from GRS~1915+105, the column density in \avd has remained at the Galactic value, and thus in our case the decrease in luminosity should be intrinsic and be related to accretion activity. 

Rapid changes in luminosity from stellar-mass to supermassive BHs have been attributed to radiation pressure instability as one of the explanations, e.g. the BHXRBs GRS~1915+105 \citep{Belloni2000,Neilsen2012} and IGR~J17091--3624 \citep{Altamirano2011,Wang2018}, the intermediate BH HLX--1 \citep{Wu2016} and AGN IC~3599 \citep{Grupe2015}. 
\cite{Wu2016} further suggests that this type of variability may exist in different accreting BHs on timescales proportional to the BH mass.
Following the empirical relationship between the bolometric luminosity and the variability duration of different type of BHs provided by \cite{Wu2016}, the timescale of the corresponding variability of \avd with a mass of $\sim 10^{6.3}\,M_{\odot}$ should be a couple of years. 
Although we missed the rising phase of the X-ray flare, the duration of the main flare (i.e. from phases 0 to 4, roughly 400\,days) is marginally consistent with the above relationship.
However, we have monitored \avd for another year after phase 4 but did not observe re-brightening in from optical to X-rays despite the sparse data, which conflicts with the scenario of recurrent flares caused by radiation pressure instability.

In jetted TDEs, \cite{Zauderer2013} argues that the rapid decline in X-rays of Swift~1644+57 corresponds to the closure of the relativistic jet, which is most likely a consequence of the decrease in mass accretion rate below the critical value. 
Regarding \avd, although there are only four radio detections spanning roughly 3\,yrs, \cite{Wang2023} was able to fit the multi-epoch radio SED with the self-absorbed synchrotron model developed by \cite{Granot2002}, and the result suggests that the radio flux increases by 50 per cent during the flaring episode (from Day~-106 to 348). 
The latest VLASS detection of \avd in Day~846 reveals a further increase in flux density at 3\,GHz to 2.8\,mJy. This is roughly six times higher than the prediction of the synchrotron model, suggesting that the radio flare is either still in the rising phase or has reached its peak in phase 5 and declines now. Although the uncertainties on the radio-band behavior of the source are large, the combination of several facts (the sub-Eddington state transition, the consequent X-ray spectral hardening, the high radio brightness of $\sim 5 \times$10$^{6}$\,K, the optically-thick radio emission at low frequencies and the variable radio luminosity of the source) reconcile with the interpretation of an accretion-ejection coupling similar to what is observed in BHXRBs \citep{fender04} and low-luminosity jetted AGN \citep{ho08,heckman14,baldi21b}, powered by an advection-dominated accretion flow (ADAF, e.g. \citealt{Narayan1994,Esin1997,falcke04}). These are evidence that jet-related activities could secondarily contribute to the X-ray emission in \avd.



If the inner region of the accretion disk evaporates into a thick ADAF as the mass accretion rate decreases, this could result in a gradually hardening X-ray spectrum. 
To interpret the fast formation of the accretion disk in \avd, \cite{Wang2023} proposes a slim disk with a large height-to-radius ratio to reduce the viscous time-scale.
If the late-time X-ray emission in \avd could be attributed to an inefficiently radiated ADAF due to a decrease in mass accretion rate, a slim disk-to-ADAF evolution is required. 
Moreover, as indicated by \cite{Wang2023}, distinguishing between a thermal or non-thermal source for the late-time X-ray emission~(Days 373--615) poses a statistical challenge. Nevertheless, it is worth noting that both the temperature and photospheric radius obtained from the \texttt{bbody} model are notably smaller than those observed during the rest time. This unphysical combination suggests that the X-ray emission may have likely departed from its thermal origin, coinciding with the VLBA detection. 
A BHXRB could be transformed from a very high luminous state to a hard dim state in less than 300\,days, while the mass accretion rate decreases by several orders of magnitude (see Fig.~12 of \citealt{Esin1997}). 
This timescale is in line with what we observed in \avd. Theoretical work is required to examine the slim-to-ADAF transition scenario.


\subsection{State transition}
Sub-Eddington systems (e.g. BHXRBs and CLAGN) undergo state transitions, either from hard to soft or vice versa. Especially, the soft to hard always occurs at a few percent of Eddington luminosity, independently of the mass of the accretor \citep{Maccarone2003,Done2007,Noda2018}. 
At the same time, state transitions have also been observed in super-Eddington systems such as ULXs \citep{Sutton2013,Middleton2015,Gurpide2021}, related to changes in mass-transfer rates and subsequent narrowing of the opening angle of the super-critical funnel, which has been proposed to play a decisive role in their spectral state (e.g. \citealt{Middleton2015,Gurpide2021}).

As discussed above, \avd has shown some similarities with BHXRBs. However, its high variability and ultra-soft spectrum make it different from canonical BHXRBs but more consistent with ULXs. Especially, its spectrum in phases 1--3 is alike to that of the supersoft ultraluminous state in ULXs (see Fig.~2 in \citealt{Kaaret2017}).
For instance, we found that \avd shares certain resemblances with the supersoft ULX in NGC~247 (e.g. \citealt{Feng2016,Alston2021,Dai2021}). 
Both sources display a spectrum primarily dominated by a blackbody, accompanied by extreme variability. Moreover, the blackbody temperature of NGC~247 ULX also remains roughly constant while its luminosity changes by a factor of $\sim6$ (see Table 3 of \citealt{Dai2021}). 
However, as \avd evolved to phases 4 and 5, its spectra turned to be a powerlaw-like spectrum, which has not been observed in supersoft ULXs. 
Although based on the adopted mass of $10^{6.3}\,M_{\odot}$ \citep{Malyali2021} the X-ray luminosities of \avd do not exceed the super-Eddington luminosity, the mass is estimated via an empirical mass-estimation technique, which carries large systematic uncertainties. We therefore do not exclude the possibility that \avd first exhibited super-Eddington accretion properties in phases 1--3 when $L_{\rm X} > 0.01\,L_{\rm Edd}$ and then switched to sub-Eddington accretion regime in phases 4--5 when $L_{\rm X} \leq 0.01\,L_{\rm Edd}$, as proposed by \cite{Wang2023}.

{\avd} has also exhibited several other characteristics that set it apart from BHXRBs, AGNs, and ULXs. Firstly, there is almost no hard X-ray emission above 2\,keV during the flare, which has been a common feature of TDEs. Besides, comparable spectral hardening behavior has been observed in some X-ray TDEs, e.g. NGC~5905 \citep{Bade1996}, RX~J1242--1119 \citep{Komossa2004} and AT2018fyk \citep{Wevers2021}, and has been argued to be evidence for the formation of a corona. 
\cite{Wang2023} reported that the late-time VLBA detection of \avd could be due to the formation of a compact jet. As hard X-rays can also be produced by relativistic particles via synchrotron radiation, it is unclear whether the spectrum hardening of \avd is due to the formation of corona, or the launch of jets, or ADAF, or the combination of several causes.
Secondly, the softer-when-brighter relationship in {\avd} is scaled with but not limited by the Eddington luminosity. Unlike some BHXRBs and AGN where this trend becomes invalid or even opposite when $L_{\rm X} > 0.02\,L_{\rm Edd}$ (e.g. \citealt{Kubota2004,Sobolewska2011}), such relationship in {\avd} holds instead for luminosities of 0.007--0.24\,$L_{\rm Edd}$ (see Fig.~\ref{fig:lumi_spec}). 
Finally, the photon index of {\avd} is very steep. Even after the source went into a \texttt{powerlaw}-dominated state, the photon index, $\Gamma$=3.1$-4.9$ in phase 4 and $\Gamma$=1.9$-2.7$ in phase 5, is still steeper than the one measured in a similar state of BHXRBs and AGN, but again consistent with a TDE scenario. We plan to continually monitor this target and see whether it will eventually evolve to the standard hard state and how long it will take to draw a full evolution of the accretion process.

\section{Conclusion} \label{sec:conclu}
{\avd} has exhibited high X-ray variability on both short (hundreds to thousands of seconds) and long (years) timescales. Together with its spectral features, it has shown some common and unique properties:

\begin{itemize}
\item{A rapid drop in X-rays occurs $\sim 225$\,days after the peak of the flare, followed by a soft-to-hard transition when the luminosity decreases down to $0.01\,L_{\rm Edd}$, and by the possible ejection of a optically-thick radio outflow \citep{Wang2023}};

\item{The softer-when-brighter relation has been observed throughout the flare: the spectrum hardens as the luminosity decreases;}

\item{The fractional rms amplitude is high with an average of 43\% and its evolution is related to spectral state; The variability may be attributed to some clumpy outflows intercepting with the X-ray emission from the accretion disk;}

\item{A soft excess has been detected at least in the relatively soft state, whose temperature remains more or less constant while the luminosity decreases by over one order of magnitude. None of the standard accretion disk model or the optically thick Comptonization or the reflection of the disk emission could explain its origin.}

\end{itemize}

\section*{Acknowledgements} 
The authors thank the anonymous referee for the constructive comments. 
The authors thank Erlin Qiao, Ian McHardy, Chris Done, Weimin Yuan, Lian Tao, and Rongfeng Shen for the discussion. 
YW acknowledges support from the Strategic Priority Research Program of the Chinese Academy of Sciences (Grant No. XDB0550200) and the Royal Society Newton Fund. 
LJ acknowledges the National Natural Science Foundation of China (Grant No. 12173103). 
RDB acknowledges the support from PRIN INAF 1.05.01.88.06 `Towards the SKA and CTA era: discovery, localisation, and physics of transient sources'.

\section*{DATA AVAILABILITY}
\sw, \Ni, {\it Chandra}. This paper employs a list of Chandra datasets, obtained by the Chandra X-ray Observatory, contained in~\dataset[DOI: cdc.170]{https://doi.org/10.25574/cdc.170}.

\appendix
\counterwithin{figure}{section}
\section{Estimation of the contamination level in the NICER FoV}\label{sec:app_contamination}
\begin{figure*} 
\centering  
\includegraphics[width=0.6\linewidth]{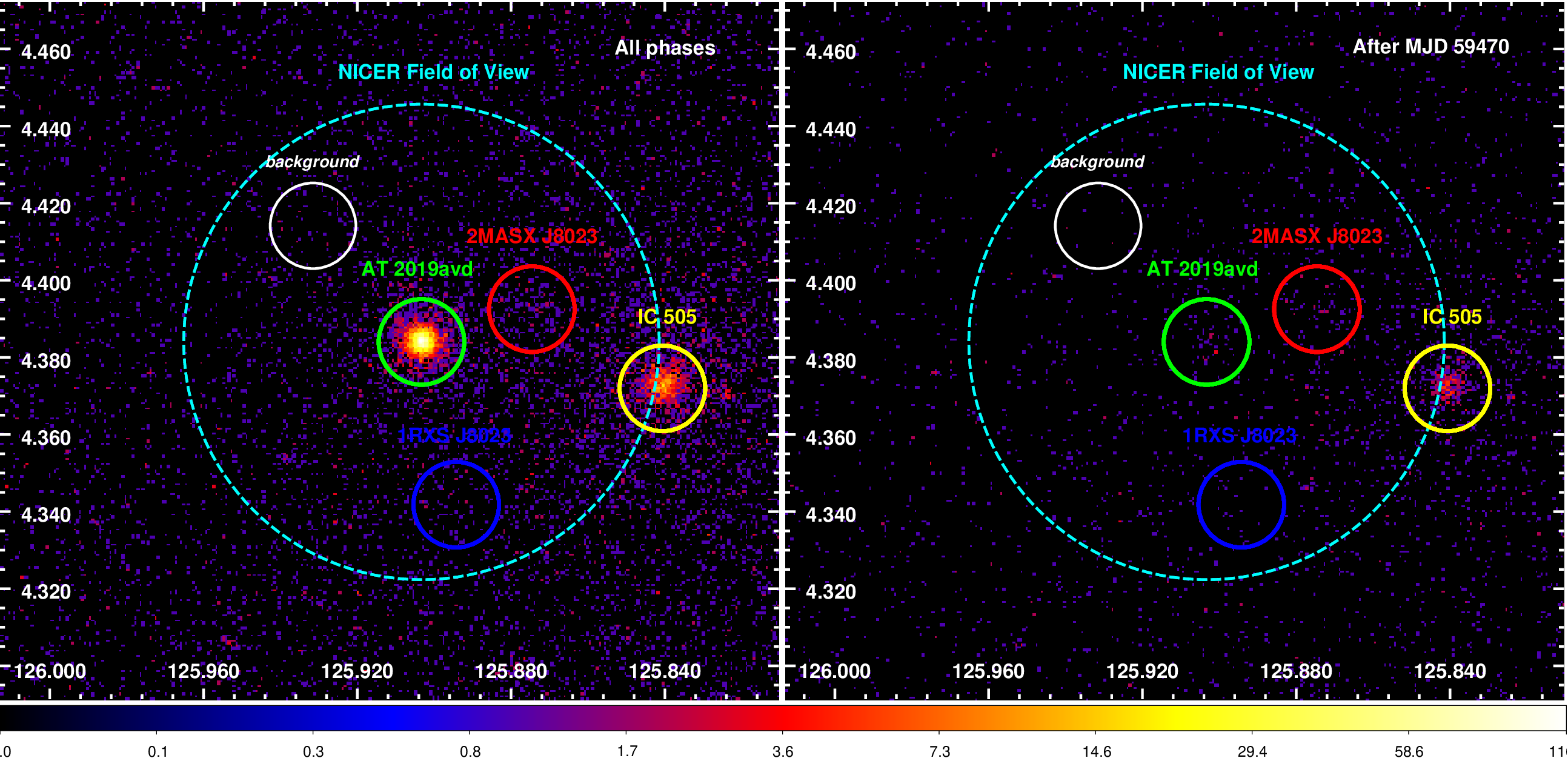}
\includegraphics[width=0.355\linewidth]{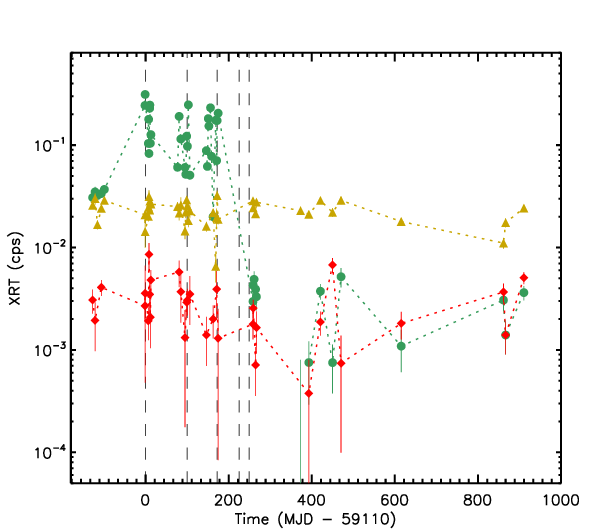}
\caption{\textbf{Left:} Stacked images of the XRT data from all the phases and after MJD~59470, respectively. The solid and dashed circulars' radii are 40\arcsec and 3.7\arcmin, respectively. The coordinates of each target are adopted from SIMBAD. \textbf{Right:} XRT lightcurves of \avd (green), IC~505 (yellow) and 2MASX~J0823 (red), calculated in the 0.3--2\,keV.}
\label{figA:app_image}
\end{figure*}

According to the record from SIMBAD, besides \avd, there were another three targets, 1RXS~J082334.6+042030, 2MASX~J08232985+0423327 and IC~505, in the NICER FoV when pointing at the location of \avd. We show the stacked images of all the XRT observations in the left panel of Fig.~\ref{figA:app_image} with a total exposure of $\sim$66\,ks. The four targets are marked with a solid circle with a radius of 40\arcsec~ in Fig.~\ref{figA:app_image} with a comparison of the NICER FoV with a radius of 3.7\arcmin. As the brightness of 1RXS~J082334.6+042030, hereafter 1RXS~J0823, is consistent with the background emission, we excluded it from the following analysis.
2MASX~J08232985+0423327, hereafter 2MASX~J0823, has been defined as an AGN, which has been marginally detected by XRT. 
The third target, IC~505, has been classified as a LINER AGN with an emission line centred around 1\,keV. 
We show the lightcurves of the three X-ray sources with XRT data in the right panel of Fig.~\ref{figA:app_image} and their best-fitting parameters inferred from different instruments in Table~\ref{tabA:para}. 
Both the lightcurve and the spectral parameters of IC~505 suggest that it is a rather stable target. \avd used to be the brightest one among the four targets when it was in the flaring episode; it evolved to be fainter than IC~505 since phase 5, which flux became comparable to 2MASX~J0823 after MJD 59470. 
The fluxes of \avd and IC~505 detected by XTI are higher than those detected by XRT, indicating that the \Ni detection have been contaminated to some extend. 
We aim to study the \Ni data from phases 1--4 when 2MASX~J0823 was still fainter than \avd. Hence, the contamination from 2MASX~J0823 should be negligible.

In the next step, we carefully examined the contamination level from IC~505 in the \Ni observations of \avd. As shown in Table~\ref{tabA:para}, there are no targets other than IC~505 presenting significant emission lines around 1\,keV, which had already been observed in the XMM-Newton spectrum of IC~505 in 2015. 
We jointly fitted the most recent \Ni data of \avd and IC~505 and found that the line flux in the former spectrum is roughly 40\% of that in the latter spectrum. 
We then defined a model, i.e. \texttt{const*tbabs*zashift(powerlaw+gaussian)} with the best-fitting parameters to the IC~505 XTI spectrum (see Table~\ref{tabA:para}) and \texttt{const}=0.4, as a template of the contamination. Finally, we loaded the template as part of the \avd data and fitted the left residuals as the net emission from \avd.

\begin{figure} 
\centering  
\includegraphics[width=0.4\linewidth]{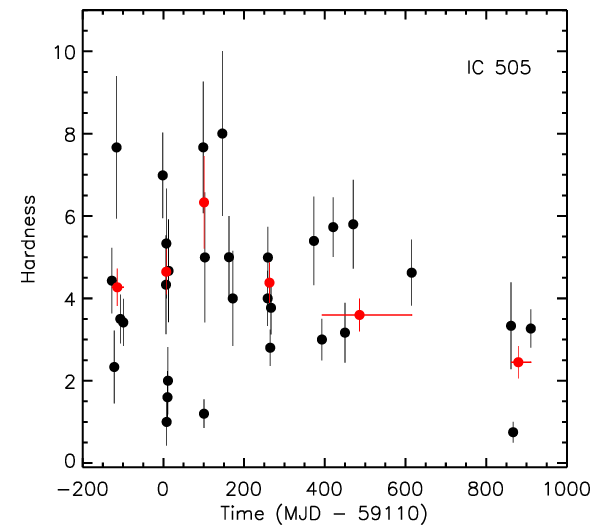}
\hspace{-0.3cm}
\includegraphics[width=0.4\linewidth]{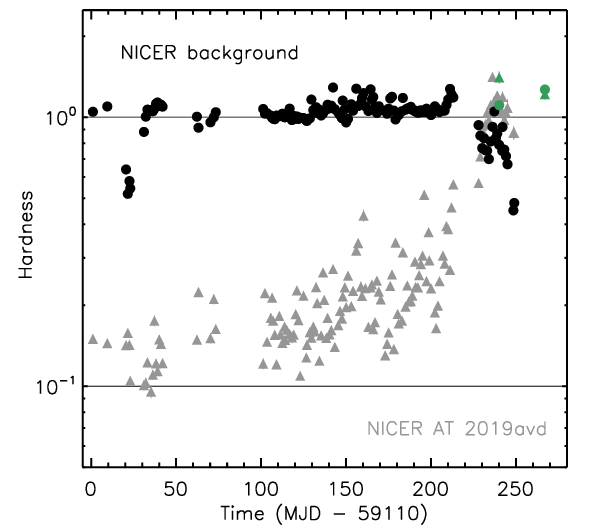}
\caption{\textbf{Left:} The hardness ratio of the XRT data of IC~505. The red dots represent the average of the hardness ratio over different periods. The periods are defined the same as in Fig.~\ref{fig:lc}. \textbf{Right:} The hardness ratios of the \Ni data (background included, grey/green triangles) and the background (black/green dots) of \avd, respectively. The green ones indicate the hardness ratio measured with the \Ni data taken on Days 855 and 882, respectively.}
\label{figA:app_hdn}
\end{figure}

To justify the spectral hardening of \avd shown in the \Ni data, we have also calculated the hardness ratio of IC~505 and the background of \avd with the XRT and the XTI data, respectively and plotted the results in Fig.~\ref{figA:app_hdn}. The figures show that the spectrum of IC~505 softened over the course of the flare of \avd. On the other hand, while the hardness of the \avd background remained roughly constant in phases 1--3 and decreased in phase 4, the background-included spectrum of \avd hardened monotonically. This fact indicates that the spectral hardening of \avd is more significant than the softening of its background. All these results support that the spectral hardening detected by \avd with the \Ni data is not an artefact.

\begin{table*}
\caption{The best-fitting parameters for the spectra of IC~505, 2MASX~J0823 and \avd}
\renewcommand{\arraystretch}{1.3}
\setlength{\tabcolsep}{3pt}
\footnotesize
\centering
\begin{tabular}{lcccccc}
\hline \hline
model   &  \begin{tabular}{@{}c@{}} IC~505  \\ EPIC-pn$^{  \forall  }$ \end{tabular}  &  \begin{tabular}{@{}c@{}} IC~505  \\ XRT \end{tabular} &  \begin{tabular}{@{}c@{}} IC~505  \\ XTI \end{tabular} &  \begin{tabular}{@{}c@{}} 2MASX~J0823  \\ XRT \end{tabular}    &\begin{tabular}{@{}c@{}} \avd  \\ XRT \end{tabular}   &\begin{tabular}{@{}c@{}} \avd \\ XTI \end{tabular}   \\
\hline
$N_{\rm H}~(\rm 10^{20}\,\rm cm^{-2})$&2.4$^{ \natural }$&2.4$^{ \natural }$& 2.4$^{ \natural }$ & 2.4$^{ \natural }$ & 2.4$^{ \natural }$ & 2.4$^{ \natural }$ \\
\hline
$\Gamma$& 1.6$\pm$0.1 & 1.3$\pm$0.4 & 1.6$\pm$0.1 & 0.9$_{-0.5}^{+1.0}$ & 2.5$\pm$0.4 & 1.7$\pm$0.1 \\
$F~(10^{-12}~\rm erg\,cm^{-2}\,s^{-1})$& 0.32$\pm$0.02 & 0.25$\pm$0.05 &1.38$\pm$0.04  & 0.08$\pm$0.02 & 0.13$\pm$0.03 & 0.89$\pm$0.05\\
\hline
$E_{\rm gau}$~(keV)& 0.93$\pm$0.01 & 0.93$\pm$0.02 & 0.96$\pm$0.01&-&1.03$^{  \ell }$ &1.03$\pm$0.03\\
eqw~(keV)& 1.41$\pm$0.17 & 0.15$\pm$0.03 & 0.52$\pm$0.04&-&0.28$^{  \ell }$ &0.28$\pm$0.07\\
$F_{\rm gau}~(10^{-12}\,\rm erg\,cm^{-2}\,s^{-1})$& 0.28$\pm$0.02 & 0.25$\pm$0.04 & 0.43$\pm$0.02 &-&$<$0.02 & 0.14$\pm$0.03\\
\hline
$F_{\rm tot}~(10^{-12}\,\rm erg\,cm^{-2}\,s^{-1})$& 0.61$\pm$0.02 & 0.50$\pm$0.03 & 1.80$\pm$0.03 &0.08$\pm$0.02 & 0.13$\pm$0.03 & 1.03$\pm$0.04\\
$\chi^2/\nu$& 21.91/24 & 84.55/117 & 106.98/85& 11.76/12 & 10.82/13 &29.68/20\\
\hline
MJD &57120&58982-60020&59965-59996&59470-60020&59470-60020& 59965\\
\hline
\end{tabular}
\begin{flushleft}
\begin{tablenotes}
 \item $ \forall $: The EPIC-pn spectrum was downloaded directly from the 4XMM-DR12 catalogue: \url{http://xmm-catalog.irap.omp.eu/}.
 \item $\natural $: The extinction is fixed at its Galactic value obtained from \cite{HI4PI}. 
 \item $\ell$: The {\sc gaussian} centroid energy and equivalent width are fixed at the best-fitting parameters obtained from the fit to the \Ni spectrum.
\label{tabA:para}
\end{tablenotes}
\end{flushleft}   
\end{table*}

\begin{figure} 
\centering  
\includegraphics[width=0.4\linewidth]{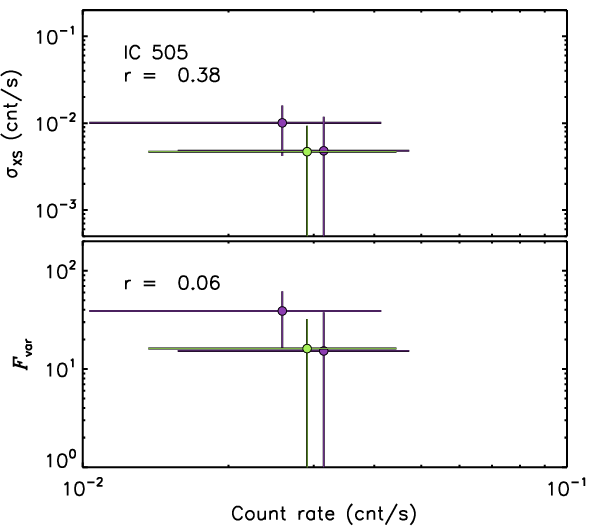}
\includegraphics[width=0.4\linewidth]{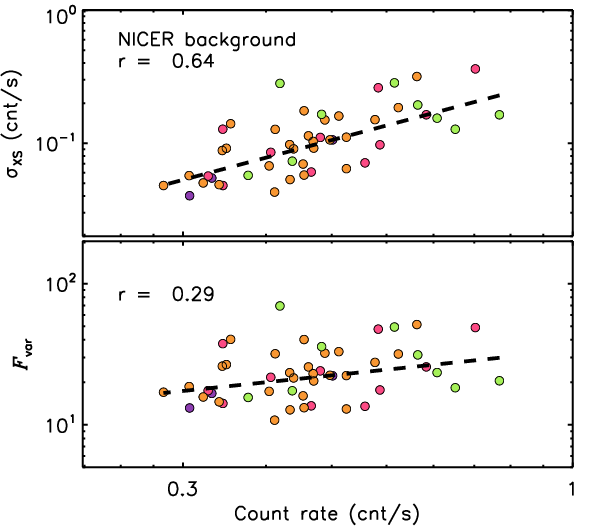}
\caption{Absolute and fractional rms vs count rate. The upper and lower panels correspond to the XRT data of IC~505 and the NICER background of \avd. $r$ represents linear Pearson correlation coefficient.
The colours are defined the same as in Fig.~\ref{fig:rms}. }
\label{figA:app_corr}
\end{figure}

\section{Measurement of fractional rms variability amplitude}\label{sec:app_rms}
We computed the fractional rms variability amplitude, $F_{\rm var}$, as described in \cite{Vaughan2003} as follows. The observed variance, $S$, can be measured from the lightcurve directly as
\begin{equation}
S^2=\frac{1}{N-1} \sum_{i=1}^N (x_{i}-\overline{x})^2,
\end{equation}
where $\overline{x}$ is the mean of the time series $x_{i}$ ($i=0,1,2,..,N$). Generally, $x_{i}$ should be evenly sampled. In the case of \avd, the variability is in the timescales longer than several hundred of seconds, which corresponds to the length of one GTI. Therefore, we defined $x_{i}$ as the average count rate of one GTI and chose the observations with two constrains, $N\geq 4$ and the observation time between $x_{0}$ and $x_{N}$ is longer than $10$\,ks, to reduce the bias in the results caused by observation cadences.

As a lightcurve $x_{i}$ should have finite uncertainties $\sigma_{{\rm err},i}$, due to the measurement errors (e.g. Poisson noise), the intrinsic source variance would correspond to the `excess variance', which is the variance after subtracting the contribution expected from the measurement errors,
\begin{equation}\label{equ:excess}
\sigma_{\rm XS}^2=S^2-\overline{\sigma_{\rm err}^2},
\end{equation}
where $\overline{\sigma_{\rm err}^2}$ is the mean square error
\begin{equation}
\overline{\sigma_{\rm err}^2}=\frac{1}{N} \sum_{i=1}^N \sigma^2_{{\rm err},i}.
\end{equation}
Here $\sigma_{\rm XS}$ is the absolute rms. $F_{\rm var}$ is the square root of the normalised excess variance,
\begin{equation}
F_{\rm var}=\sqrt{\frac{S^2-\overline{\sigma_{\rm err}^2}}{\overline{x}^2}}.
\end{equation}
Finally, the uncertainty on $F_{\rm var}$ is given by
\begin{equation}
{\rm err}(F_{\rm var})=\sqrt{(\frac{1}{2N}\frac{\overline{\sigma_{\rm err}^2}}{\overline{x}^2 F_{\rm var}})^2+(\frac{\overline{\sigma_{\rm err}^2}}{N}\frac{1}{\overline{x}})^2}.
\end{equation}

We have also calculated $\sigma_{XS}$ and $F_{\rm var}$ of IC~505 and the \avd background with the XRT and the XTI data, and their Pearson correlation. 
Unfortunately, there are only a few XRT data of IC~505 fulfilling the criteria for the computation of $F_{\rm var}$, and the most significant detection is $F_{\rm var}=0.39\pm0.23$. Regarding the background, its absolute rms amplitude is strongly correlated to the count rate ($r>0.5$); the fractional rms amplitude is also correlated to the count rate but the dependence reduces ($r<0.3$). Overall, the high variability of  \avd detected by \Ni should be entirely intrinsic.

\section{Temporal evolution of the spectral parameters}\label{sec:app_spec}
We illustrate the evolution of the spectral parameters along with the flare in Fig.~\ref{fig:para_pl}.
As shown in the first panel, the column density ranges among (0.24--0.86)$\times 10^{21}\,\rm cm^{-2}$ and in some observations, it pegs at the lower limit which is the value of the Galactic X-ray absorption in the direction of {\avd} \citep{HI4PI}. 
We also see from the figure that while the \texttt{bbody} flux changes over one order of magnitude from phase 1 to phase 3 (the fourth panel), the \texttt{bbody} temperature remains more or less constant although accompanied by some scatters (the second panel). 
The \texttt{bbody} temperatures obtained by fitting the XRT spectra are added to the same panel of Fig.~\ref{fig:para_pl} as magenta dots which are in good agreement with the {\Ni} data.
The photon index ranges between 1.22 and 6.23 and tends to disperse. When the flux significantly dropped in phase 4 (the fifth panel), $\Gamma$ is on average lower than other phases, indicating the hardening of the spectrum.
\begin{figure} 
\centering  
\resizebox{0.5\columnwidth}{!}{\rotatebox{270}{\includegraphics[clip]{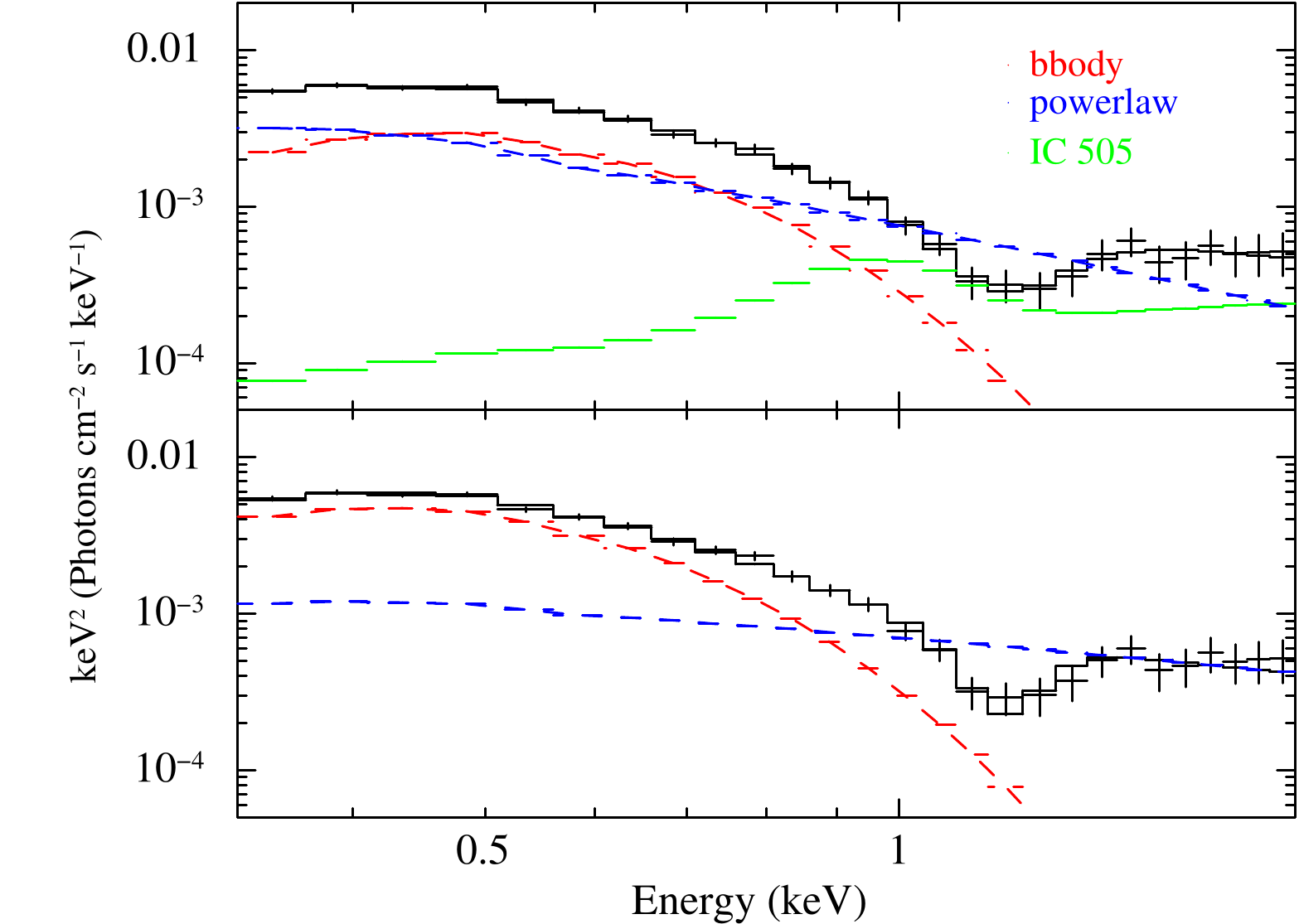}}}
\caption{Unfolded spectra of ObsID. 3201770178 from phase 2 with (upper) and without (lower) the subtraction of the IC~505 template.}
\label{fig:abs_compare}
\end{figure}

\begin{figure}
\centering      \mbox{\includegraphics[angle=0,width=0.5\linewidth]{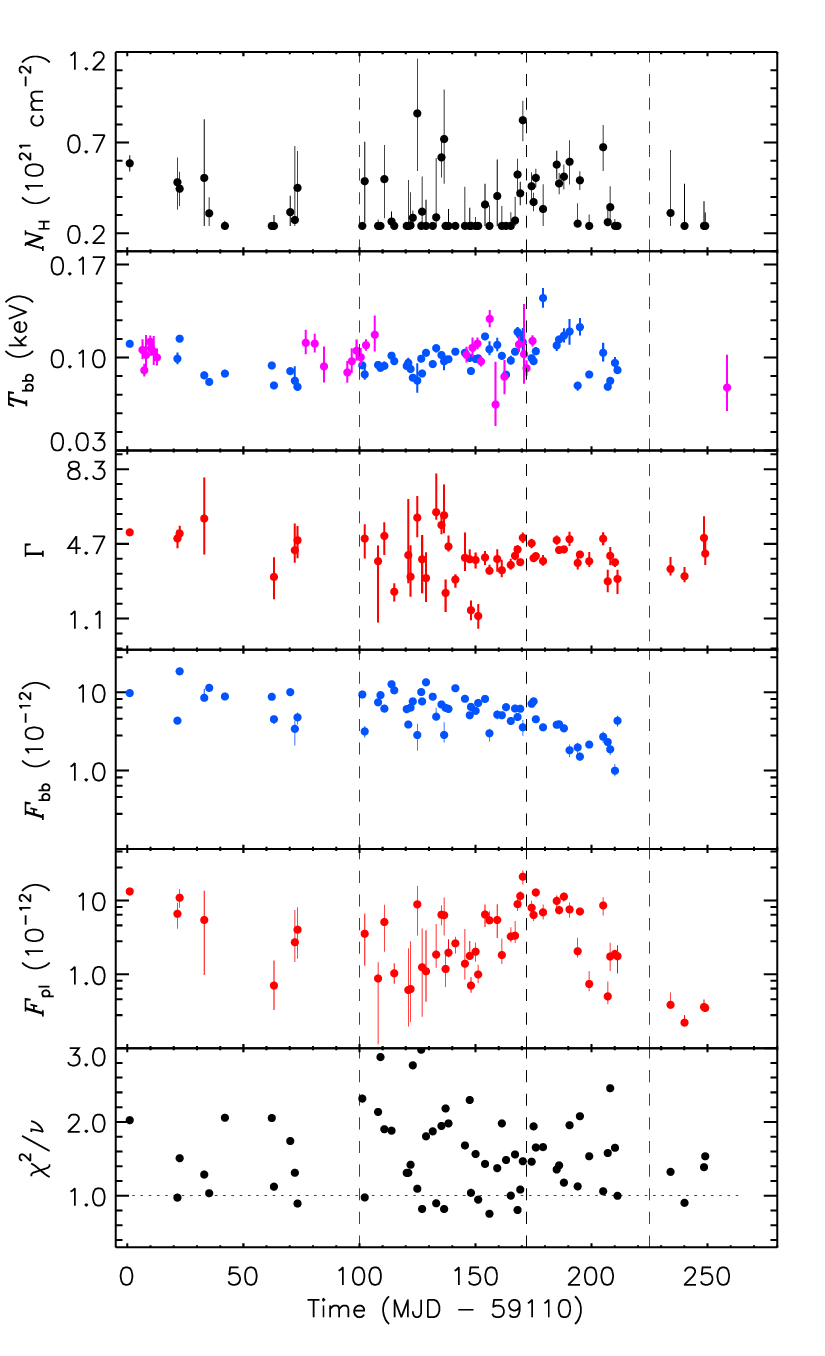}}
\vspace{-0.2cm}
\caption{Evolution of the spectral parameters. From top to bottom, they are the column density, the blackbody temperature, the photon index, the \texttt{bbody} flux, the \texttt{powerlaw} flux and the reduced $\chi^2$. 
The magenta dots represent the values of $T_{\rm bb}$ inferred from the XRT spectrum.}
\label{fig:para_pl}
\end{figure}

\section{Determination of the Power Spectral Densities and Search for Periodicity}\label{sec:app_ls}
To search for any periodicity of the \Ni data, we conducted an analysis of the power spectral densities (PSD) of the \Ni long-term lightcurve in units of GTI. Briefly, the method we used is an adaption of the original method proposed by \cite{Done1992} and improved by \cite{Uttley2002} which relies on simulating lightcurves with a given PSD and imprinting the same sampling pattern as the observed one, in order to reverse engineer the process that generated the observed lightcurve. The lightcurves were simulated using the method of \cite{Emmanoulopoulos2013}, initially with a binning equal to 1/2 of the minimum exposure time of the \Ni GTI and 20 times longer than the observed lightcurve, to introduce aliasing and red-noise leakage effects. 
For the PSD, we chose a \texttt{powerlaw} model ($S \propto f^-\beta$) characteristic of accreting sources \citep{Vaughan2003}, whereas for the probability density function we fitted the observed count rates using two log normal distributions. Because in practice we are simulated a long lightcurve and snipping it into segments of the desired length, we additionally added a break at $0.5/T$ to the PSD following \cite{Middleton2010}, after which the PSD breaks to a flat powerlaw. 

We generated 2,000 lightcurves for each trial $\beta$ value ranging from 0.5 to 3.2 in steps of 0.1, added Poisson noise based on the lightcurve exposures times and compared the Lomb-Scargle periodogram of the observed data with that of the simulations using the statistic proposed by \cite{Uttley2002}. 
We determined rejection probabilities following their work and uncertainties by adding 1$\sigma$ to the rejection probability found for the best-fit value following \cite{Markowitz2010}. We are aware that determining uncertainties using the method of \cite{Uttley2002} is a subtle issue that likely requires additional Monte Carlo simulations (see discussion in \citet{Markowitz2010}) but in order to limit the computational burden we assume the simplistic approach introduced by \citet{Markowitz2010}. We also note that considering the uncertainties as defined by \cite{Uttley2002} we obtain very similar results (see below). 

\begin{figure}
    \centering
    \includegraphics[width=0.4\textwidth]{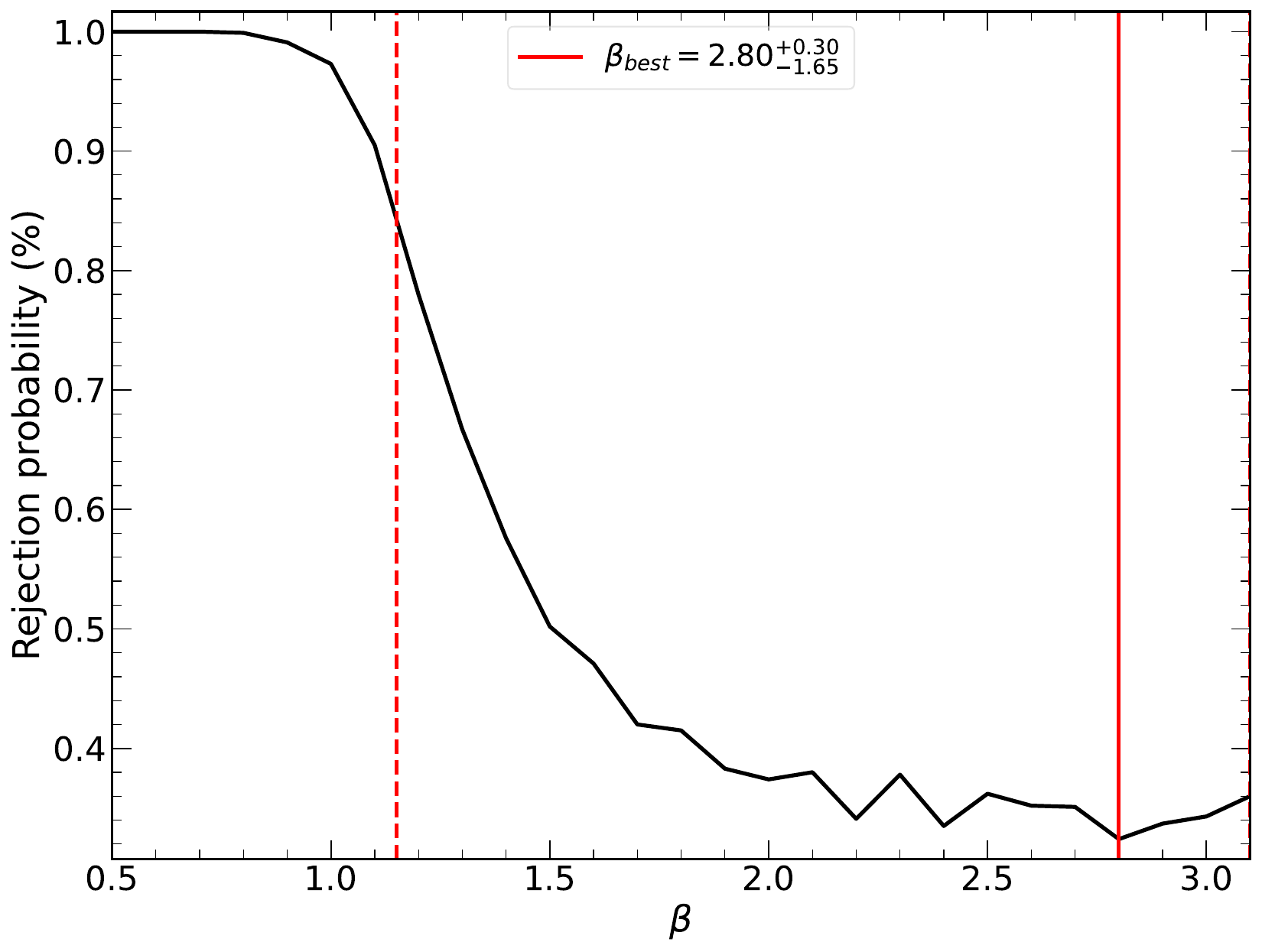}
    \caption{Rejection probability contour for the index of the powerlaw PSD model ($S \propto f^-\beta$). The best-fit value is indicated by a solid line and the approximate uncertainties (see text for details) by a dashed line.}
    \label{fig:beta_contour}
\end{figure}

We evaluated the periodogram from 1/2$<dt>$, where $<dt>$ is the median sampling time (a replacement for the nonexistent Nyquist frequency \citep{VanderPlas2018}), to 1/$T$ where $T$ was the baseline of the GTIs, using a frequency spacing equal to $N\times1/T$ where $N$ is the oversampling factor which we set to 2 (which we justify below). The periodograms were rebinned logarithmically following \cite{Papadakis1993} -- so less samples are required to reach Gaussianity -- to have at least 20 powers/bin. We note that while the powers in the Lomb-Scargle periodogram are known \textit{not} to be independent -- violating the assumption of independent samples inherent in the $\chi^2$ statistic -- we found this problem is mitigated as long as the periodogram is rebinned and not too oversampled. We verified this through Monte Carlo simulations of lightcurves with known PSD, and found that with a logarithmic rebinning of $\sim$15--20 powers/bin and an oversample factor of 1--2 the recovered $\beta$ value had a bias of 0.1 or less when using $\chi^2$ statistic. The resulting contours for $\beta$ are shown in Figure~\ref{fig:beta_contour}. This simple model cannot be statistically rejected, and a rather wide range of $\beta$ values would be compatible with the data (rejection probability $<$60\%). Approximately our best-fit beta value is found to be $\beta=2.8_{-1.65}^{+0.3}$. Considering the uncertainties as defined by \cite{Uttley2002} would reduce the lower limit only slightly, to $\sim$1.3.

We next used the best-fit model (together with its uncertainties) as the null hypothesis to test for the presence of any periodicity in the periodogram. We retrieved 5,000 lightcurves taking into account the uncertainties on $\beta$ (so as to take the uncertainties into account for the false alarm probability derivation \citep{Vaughan2005}) and looked for the highest peak in the corresponding periodograms of each generated lightcurve. The results are shown in Fig.~\ref{fig:lomb-scargle} which show the 3$\sigma$ one- (green line) and multiple-(dashed line) trial(s) false alarm probabilities. We see that there is no obvious peak above the false alarm probability levels. The highest peak at around 25\,days is well below the one-trial false alarm probability. We thus conclude that the variability of \avd is fully consistent with the typical aperiodic variability commonly observed across all accreting systems.

\bibliography{cite}{}

\begin{thebibliography}{}
\expandafter\ifx\csname natexlab\endcsname\relax\def\natexlab#1{#1}\fi
\providecommand{\url}[1]{\href{#1}{#1}}
\providecommand{\dodoi}[1]{doi:~\href{http://doi.org/#1}{\nolinkurl{#1}}}
\providecommand{\doeprint}[1]{\href{http://ascl.net/#1}{\nolinkurl{http://ascl.net/#1}}}
\providecommand{\doarXiv}[1]{\href{https://arxiv.org/abs/#1}{\nolinkurl{https://arxiv.org/abs/#1}}}

\bibitem[{{Alabarta} {et~al.}(2022){Alabarta}, {M{\'e}ndez}, {Garc{\'\i}a},
  {Peirano}, {Altamirano}, {Zhang}, \& {Karpouzas}}]{Alabarta2022}
{Alabarta}, K., {M{\'e}ndez}, M., {Garc{\'\i}a}, F., {et~al.} 2022, \mnras,
  514, 2839, \dodoi{10.1093/mnras/stac1533}

\bibitem[{{Alexander} {et~al.}(2020){Alexander}, {van Velzen}, {Horesh}, \&
  {Zauderer}}]{Alexander2020}
{Alexander}, K.~D., {van Velzen}, S., {Horesh}, A., \& {Zauderer}, B.~A. 2020,
  \ssr, 216, 81, \dodoi{10.1007/s11214-020-00702-w}

\bibitem[{{Alston} {et~al.}(2021){Alston}, {Pinto}, {Barret}, {D'A{\`\i}}, {Del
  Santo}, {Earnshaw}, {Fabian}, {Fuerst}, {Kara}, {Kosec}, {Middleton},
  {Parker}, {Pintore}, {Robba}, {Roberts}, {Sathyaprakash}, {Walton}, \&
  {Ambrosi}}]{Alston2021}
{Alston}, W.~N., {Pinto}, C., {Barret}, D., {et~al.} 2021, \mnras, 505, 3722,
  \dodoi{10.1093/mnras/stab1473}

\bibitem[{{Altamirano} \& {Strohmayer}(2012)}]{Altamirano2012}
{Altamirano}, D., \& {Strohmayer}, T. 2012, \apjl, 754, L23,
  \dodoi{10.1088/2041-8205/754/2/L23}

\bibitem[{{Altamirano} {et~al.}(2011){Altamirano}, {Belloni}, {Linares}, {van
  der Klis}, {Wijnands}, {Curran}, {Kalamkar}, {Stiele}, {Motta},
  {Mu{\~n}oz-Darias}, {Casella}, \& {Krimm}}]{Altamirano2011}
{Altamirano}, D., {Belloni}, T., {Linares}, M., {et~al.} 2011, \apjl, 742, L17,
  \dodoi{10.1088/2041-8205/742/2/L17}

\bibitem[{{Atapin} {et~al.}(2019){Atapin}, {Fabrika}, \&
  {Caballero-Garc{\'\i}a}}]{Atapin2019}
{Atapin}, K., {Fabrika}, S., \& {Caballero-Garc{\'\i}a}, M.~D. 2019, \mnras,
  486, 2766, \dodoi{10.1093/mnras/stz1027}

\bibitem[{{Auchettl} {et~al.}(2017){Auchettl}, {Guillochon}, \&
  {Ramirez-Ruiz}}]{Auchettl2017}
{Auchettl}, K., {Guillochon}, J., \& {Ramirez-Ruiz}, E. 2017, \apj, 838, 149,
  \dodoi{10.3847/1538-4357/aa633b}

\bibitem[{{Bade} {et~al.}(1996){Bade}, {Komossa}, \& {Dahlem}}]{Bade1996}
{Bade}, N., {Komossa}, S., \& {Dahlem}, M. 1996, \aap, 309, L35

\bibitem[{{Baldi} {et~al.}(2021){Baldi}, {Williams}, {Beswick}, {McHardy},
  {Dullo}, {Knapen}, {Zanisi}, {Argo}, {Aalto}, {Alberdi}, {Baan}, {Bendo},
  {Fenech}, {Green}, {Kl{\"o}ckner}, {K{\"o}rding}, {Maccarone}, {Marcaide},
  {Mutie}, {Panessa}, {P{\'e}rez-Torres}, {Romero-Ca{\~n}izales}, {Saikia},
  {Saikia}, {Shankar}, {Spencer}, {Stevens}, {Uttley}, {Brinks}, {Corbel},
  {Mart{\'\i}-Vidal}, {Mundell}, {Pahari}, \& {Ward}}]{baldi21b}
{Baldi}, R.~D., {Williams}, D.~R.~A., {Beswick}, R.~J., {et~al.} 2021, \mnras,
  508, 2019, \dodoi{10.1093/mnras/stab2613}

\bibitem[{{Ballantyne} {et~al.}(2001){Ballantyne}, {Ross}, \&
  {Fabian}}]{Ballantyne2001}
{Ballantyne}, D.~R., {Ross}, R.~R., \& {Fabian}, A.~C. 2001, \mnras, 327, 10,
  \dodoi{10.1046/j.1365-8711.2001.04432.x}

\bibitem[{{Bellm} {et~al.}(2019){Bellm}, {Kulkarni}, {Graham}, {Dekany},
  {Smith}, {Riddle}, {Masci}, {Helou}, {Prince}, {Adams}, {Barbarino},
  {Barlow}, {Bauer}, {Beck}, {Belicki}, {Biswas}, {Blagorodnova}, {Bodewits},
  {Bolin}, {Brinnel}, {Brooke}, {Bue}, {Bulla}, {Burruss}, {Cenko}, {Chang},
  {Connolly}, {Coughlin}, {Cromer}, {Cunningham}, {De}, {Delacroix}, {Desai},
  {Duev}, {Eadie}, {Farnham}, {Feeney}, {Feindt}, {Flynn}, {Franckowiak},
  {Frederick}, {Fremling}, {Gal-Yam}, {Gezari}, {Giomi}, {Goldstein},
  {Golkhou}, {Goobar}, {Groom}, {Hacopians}, {Hale}, {Henning}, {Ho}, {Hover},
  {Howell}, {Hung}, {Huppenkothen}, {Imel}, {Ip}, {Ivezi{\'c}}, {Jackson},
  {Jones}, {Juric}, {Kasliwal}, {Kaspi}, {Kaye}, {Kelley}, {Kowalski},
  {Kramer}, {Kupfer}, {Landry}, {Laher}, {Lee}, {Lin}, {Lin}, {Lunnan},
  {Giomi}, {Mahabal}, {Mao}, {Miller}, {Monkewitz}, {Murphy}, {Ngeow},
  {Nordin}, {Nugent}, {Ofek}, {Patterson}, {Penprase}, {Porter}, {Rauch},
  {Rebbapragada}, {Reiley}, {Rigault}, {Rodriguez}, {van Roestel}, {Rusholme},
  {van Santen}, {Schulze}, {Shupe}, {Singer}, {Soumagnac}, {Stein}, {Surace},
  {Sollerman}, {Szkody}, {Taddia}, {Terek}, {Van Sistine}, {van Velzen},
  {Vestrand}, {Walters}, {Ward}, {Ye}, {Yu}, {Yan}, \& {Zolkower}}]{Bellm2019}
{Bellm}, E.~C., {Kulkarni}, S.~R., {Graham}, M.~J., {et~al.} 2019, \pasp, 131,
  018002, \dodoi{10.1088/1538-3873/aaecbe}

\bibitem[{{Belloni} {et~al.}(2005){Belloni}, {Homan}, {Casella}, {van der
  Klis}, {Nespoli}, {Lewin}, {Miller}, \& {M{\'e}ndez}}]{Belloni2005}
{Belloni}, T., {Homan}, J., {Casella}, P., {et~al.} 2005, \aap, 440, 207,
  \dodoi{10.1051/0004-6361:20042457}

\bibitem[{{Belloni} {et~al.}(2000){Belloni}, {Klein-Wolt}, {M{\'e}ndez}, {van
  der Klis}, \& {van Paradijs}}]{Belloni2000}
{Belloni}, T., {Klein-Wolt}, M., {M{\'e}ndez}, M., {van der Klis}, M., \& {van
  Paradijs}, J. 2000, \aap, 355, 271, \dodoi{10.48550/arXiv.astro-ph/0001103}

\bibitem[{{Blagorodnova} {et~al.}(2018){Blagorodnova}, {Neill}, {Walters},
  {Kulkarni}, {Fremling}, {Ben-Ami}, {Dekany}, {Fucik}, {Konidaris}, {Nash},
  {Ngeow}, {Ofek}, {O' Sullivan}, {Quimby}, {Ritter}, \&
  {Vyhmeister}}]{Blagorodnova2018}
{Blagorodnova}, N., {Neill}, J.~D., {Walters}, R., {et~al.} 2018, \pasp, 130,
  035003, \dodoi{10.1088/1538-3873/aaa53f}

\bibitem[{{Brown} {et~al.}(2017){Brown}, {Holoien}, {Auchettl}, {Stanek},
  {Kochanek}, {Shappee}, {Prieto}, \& {Grupe}}]{Brown2017}
{Brown}, J.~S., {Holoien}, T.~W.~S., {Auchettl}, K., {et~al.} 2017, \mnras,
  466, 4904, \dodoi{10.1093/mnras/stx033}

\bibitem[{{Chen} {et~al.}(2022){Chen}, {Dou}, \& {Shen}}]{Chen2022}
{Chen}, J.-H., {Dou}, L.-M., \& {Shen}, R.-F. 2022, \apj, 928, 63,
  \dodoi{10.3847/1538-4357/ac558d}

\bibitem[{{D'A{\`\i}} {et~al.}(2021){D'A{\`\i}}, {Pinto}, {Del Santo},
  {Pintore}, {Soria}, {Robba}, {Ambrosi}, {Alston}, {Barret}, {Fabian},
  {F{\"u}rst}, {Kara}, {Kosec}, {Middleton}, {Roberts}, {Rodriguez-Castillo},
  \& {Walton}}]{Dai2021}
{D'A{\`\i}}, A., {Pinto}, C., {Del Santo}, M., {et~al.} 2021, \mnras, 507,
  5567, \dodoi{10.1093/mnras/stab2427}

\bibitem[{{Done} {et~al.}(2007){Done}, {Gierli{\'n}ski}, \&
  {Kubota}}]{Done2007}
{Done}, C., {Gierli{\'n}ski}, M., \& {Kubota}, A. 2007, \aapr, 15, 1,
  \dodoi{10.1007/s00159-007-0006-1}

\bibitem[{{Done} {et~al.}(1992){Done}, {Madejski}, {Mushotzky}, {Turner},
  {Koyama}, \& {Kunieda}}]{Done1992}
{Done}, C., {Madejski}, G.~M., {Mushotzky}, R.~F., {et~al.} 1992, \apj, 400,
  138, \dodoi{10.1086/171979}

\bibitem[{{Emmanoulopoulos} {et~al.}(2013){Emmanoulopoulos}, {McHardy}, \&
  {Papadakis}}]{Emmanoulopoulos2013}
{Emmanoulopoulos}, D., {McHardy}, I.~M., \& {Papadakis}, I.~E. 2013, \mnras,
  433, 907, \dodoi{10.1093/mnras/stt764}

\bibitem[{{Esin} {et~al.}(1997){Esin}, {McClintock}, \& {Narayan}}]{Esin1997}
{Esin}, A.~A., {McClintock}, J.~E., \& {Narayan}, R. 1997, \apj, 489, 865,
  \dodoi{10.1086/304829}

\bibitem[{{Falcke} {et~al.}(2004){Falcke}, {K{\"o}rding}, \&
  {Markoff}}]{falcke04}
{Falcke}, H., {K{\"o}rding}, E., \& {Markoff}, S. 2004, \aap, 414, 895,
  \dodoi{10.1051/0004-6361:20031683}

\bibitem[{{Fender} \& {Belloni}(2004)}]{Fender2004}
{Fender}, R., \& {Belloni}, T. 2004, \araa, 42, 317,
  \dodoi{10.1146/annurev.astro.42.053102.134031}

\bibitem[{{Fender} {et~al.}(2004){Fender}, {Belloni}, \& {Gallo}}]{fender04}
{Fender}, R.~P., {Belloni}, T.~M., \& {Gallo}, E. 2004, \mnras, 355, 1105,
  \dodoi{10.1111/j.1365-2966.2004.08384.x}

\bibitem[{{Feng} {et~al.}(2016){Feng}, {Tao}, {Kaaret}, \&
  {Gris{\'e}}}]{Feng2016}
{Feng}, H., {Tao}, L., {Kaaret}, P., \& {Gris{\'e}}, F. 2016, \apj, 831, 117,
  \dodoi{10.3847/0004-637X/831/2/117}

\bibitem[{{Gezari}(2021)}]{Gezari2021}
{Gezari}, S. 2021, \araa, 59, \dodoi{10.1146/annurev-astro-111720-030029}

\bibitem[{{Gezari} {et~al.}(2012){Gezari}, {Chornock}, {Rest}, {Huber},
  {Forster}, {Berger}, {Challis}, {Neill}, {Martin}, {Heckman}, {Lawrence},
  {Norman}, {Narayan}, {Foley}, {Marion}, {Scolnic}, {Chomiuk}, {Soderberg},
  {Smith}, {Kirshner}, {Riess}, {Smartt}, {Stubbs}, {Tonry}, {Wood-Vasey},
  {Burgett}, {Chambers}, {Grav}, {Heasley}, {Kaiser}, {Kudritzki}, {Magnier},
  {Morgan}, \& {Price}}]{Gezari2012}
{Gezari}, S., {Chornock}, R., {Rest}, A., {et~al.} 2012, \nat, 485, 217,
  \dodoi{10.1038/nature10990}

\bibitem[{{Gierli{\'n}ski} \& {Done}(2004)}]{Gierlinski2004}
{Gierli{\'n}ski}, M., \& {Done}, C. 2004, \mnras, 349, L7,
  \dodoi{10.1111/j.1365-2966.2004.07687.x}

\bibitem[{{Gleissner} {et~al.}(2004){Gleissner}, {Wilms}, {Pottschmidt},
  {Uttley}, {Nowak}, \& {Staubert}}]{Gleissner2004}
{Gleissner}, T., {Wilms}, J., {Pottschmidt}, K., {et~al.} 2004, \aap, 414,
  1091, \dodoi{10.1051/0004-6361:20031684}

\bibitem[{{Granot} \& {Sari}(2002)}]{Granot2002}
{Granot}, J., \& {Sari}, R. 2002, \apj, 568, 820, \dodoi{10.1086/338966}

\bibitem[{{Grupe} {et~al.}(2015){Grupe}, {Komossa}, \& {Saxton}}]{Grupe2015}
{Grupe}, D., {Komossa}, S., \& {Saxton}, R. 2015, \apjl, 803, L28,
  \dodoi{10.1088/2041-8205/803/2/L28}

\bibitem[{{G{\'u}rpide} {et~al.}(2021){G{\'u}rpide}, {Godet}, {Vasilopoulos},
  {Webb}, \& {Olive}}]{Gurpide2021}
{G{\'u}rpide}, A., {Godet}, O., {Vasilopoulos}, G., {Webb}, N.~A., \& {Olive},
  J.~F. 2021, \aap, 654, A10, \dodoi{10.1051/0004-6361/202140781}

\bibitem[{{Hammerstein} {et~al.}(2022){Hammerstein}, {van Velzen}, {Gezari},
  {Cenko}, {Yao}, {Ward}, {Frederick}, {Villanueva}, {Somalwar}, {Graham},
  {Kulkarni}, {Stern}, {Andreoni}, {Bellm}, {Dekany}, {Dhawan}, {Drake},
  {Fremling}, {Gatkine}, {Groom}, {Ho}, {Kasliwal}, {Karambelkar}, {Kool},
  {Masci}, {Medford}, {Perley}, {Purdum}, {van Roestel}, {Sharma}, {Sollerman},
  {Taggart}, \& {Yan}}]{Hammerstein2022}
{Hammerstein}, E., {van Velzen}, S., {Gezari}, S., {et~al.} 2022, arXiv
  e-prints, arXiv:2203.01461.
\newblock \doarXiv{2203.01461}

\bibitem[{{Heckman} \& {Best}(2014)}]{heckman14}
{Heckman}, T.~M., \& {Best}, P.~N. 2014, \araa, 52, 589,
  \dodoi{10.1146/annurev-astro-081913-035722}

\bibitem[{{Heil} \& {Vaughan}(2010)}]{Heil2010}
{Heil}, L.~M., \& {Vaughan}, S. 2010, \mnras, 405, L86,
  \dodoi{10.1111/j.1745-3933.2010.00864.x}

\bibitem[{{Heil} {et~al.}(2012){Heil}, {Vaughan}, \& {Uttley}}]{Heil2012}
{Heil}, L.~M., {Vaughan}, S., \& {Uttley}, P. 2012, \mnras, 422, 2620,
  \dodoi{10.1111/j.1365-2966.2012.20824.x}

\bibitem[{{HI4PI Collaboration} {et~al.}(2016){HI4PI Collaboration}, {Ben
  Bekhti}, {Fl{\"o}er}, {Keller}, {Kerp}, {Lenz}, {Winkel}, {Bailin},
  {Calabretta}, {Dedes}, {Ford}, {Gibson}, {Haud}, {Janowiecki}, {Kalberla},
  {Lockman}, {McClure-Griffiths}, {Murphy}, {Nakanishi}, {Pisano}, \&
  {Staveley-Smith}}]{HI4PI}
{HI4PI Collaboration}, {Ben Bekhti}, N., {Fl{\"o}er}, L., {et~al.} 2016, \aap,
  594, A116, \dodoi{10.1051/0004-6361/201629178}

\bibitem[{{Hills}(1975)}]{Hills1975}
{Hills}, J.~G. 1975, \nat, 254, 295, \dodoi{10.1038/254295a0}

\bibitem[{{Hinkle} {et~al.}(2020){Hinkle}, {Holoien}, {Shappee}, {Auchettl},
  {Kochanek}, {Stanek}, {Payne}, \& {Thompson}}]{Hinkle2020}
{Hinkle}, J.~T., {Holoien}, T. W.~S., {Shappee}, B.~J., {et~al.} 2020, \apjl,
  894, L10, \dodoi{10.3847/2041-8213/ab89a2}

\bibitem[{{Hjellming} {et~al.}(1999){Hjellming}, {Rupen}, {Mioduszewski},
  {Kuulkers}, {McCollough}, {Harmon}, {Buxton}, {Sood}, {Tzioumis}, {Rayner},
  {Dieters}, \& {Durouchoux}}]{Hjellming1999}
{Hjellming}, R.~M., {Rupen}, M.~P., {Mioduszewski}, A.~J., {et~al.} 1999, \apj,
  514, 383, \dodoi{10.1086/306948}

\bibitem[{{Ho}(2008)}]{ho08}
{Ho}, L.~C. 2008, \araa, 46, 475,
  \dodoi{10.1146/annurev.astro.45.051806.110546}

\bibitem[{{Holoien} {et~al.}(2016){Holoien}, {Kochanek}, {Prieto}, {Stanek},
  {Dong}, {Shappee}, {Grupe}, {Brown}, {Basu}, {Beacom}, {Bersier},
  {Brimacombe}, {Danilet}, {Falco}, {Guo}, {Jose}, {Herczeg}, {Long},
  {Pojmanski}, {Simonian}, {Szczygie{\l}}, {Thompson}, {Thorstensen}, {Wagner},
  \& {Wo{\'z}niak}}]{Holoien2016a}
{Holoien}, T.~W.~S., {Kochanek}, C.~S., {Prieto}, J.~L., {et~al.} 2016, \mnras,
  455, 2918, \dodoi{10.1093/mnras/stv2486}

\bibitem[{{Holoien} {et~al.}(2020){Holoien}, {Auchettl}, {Tucker}, {Shappee},
  {Patel}, {Miller-Jones}, {Mockler}, {Groenewald}, {Hinkle}, {Brown},
  {Kochanek}, {Stanek}, {Chen}, {Dong}, {Prieto}, {Thompson}, {Beaton},
  {Connor}, {Cowperthwaite}, {Dahmen}, {French}, {Morrell}, {Buckley},
  {Gromadzki}, {Roy}, {Coulter}, {Dimitriadis}, {Foley}, {Kilpatrick}, {Piro},
  {Rojas-Bravo}, {Siebert}, \& {van Velzen}}]{Holoien2020}
{Holoien}, T. W.~S., {Auchettl}, K., {Tucker}, M.~A., {et~al.} 2020, \apj, 898,
  161, \dodoi{10.3847/1538-4357/ab9f3d}

\bibitem[{{Homan} {et~al.}(2003){Homan}, {Klein-Wolt}, {Rossi}, {Miller},
  {Wijnands}, {Belloni}, {van der Klis}, \& {Lewin}}]{Homan2003}
{Homan}, J., {Klein-Wolt}, M., {Rossi}, S., {et~al.} 2003, \apj, 586, 1262,
  \dodoi{10.1086/367699}

\bibitem[{{Jiang} {et~al.}(2014){Jiang}, {Stone}, \& {Davis}}]{Jiang2014}
{Jiang}, Y.-F., {Stone}, J.~M., \& {Davis}, S.~W. 2014, \apj, 796, 106,
  \dodoi{10.1088/0004-637X/796/2/106}

\bibitem[{{Jin} {et~al.}(2021){Jin}, {Done}, \& {Ward}}]{Jin2021}
{Jin}, C., {Done}, C., \& {Ward}, M. 2021, \mnras, 500, 2475,
  \dodoi{10.1093/mnras/staa3386}

\bibitem[{{Kaaret} {et~al.}(2017){Kaaret}, {Feng}, \& {Roberts}}]{Kaaret2017}
{Kaaret}, P., {Feng}, H., \& {Roberts}, T.~P. 2017, \araa, 55, 303,
  \dodoi{10.1146/annurev-astro-091916-055259}

\bibitem[{{Kaastra} \& {Bleeker}(2016)}]{Kaastra2016}
{Kaastra}, J.~S., \& {Bleeker}, J.~A.~M. 2016, \aap, 587, A151,
  \dodoi{10.1051/0004-6361/201527395}

\bibitem[{{Kallman} \& {Bautista}(2001)}]{Kallman2001}
{Kallman}, T., \& {Bautista}, M. 2001, \apjs, 133, 221, \dodoi{10.1086/319184}

\bibitem[{{Kara} {et~al.}(2016){Kara}, {Alston}, {Fabian}, {Cackett}, {Uttley},
  {Reynolds}, \& {Zoghbi}}]{Kara2016}
{Kara}, E., {Alston}, W.~N., {Fabian}, A.~C., {et~al.} 2016, \mnras, 462, 511,
  \dodoi{10.1093/mnras/stw1695}

\bibitem[{{Kara} {et~al.}(2018){Kara}, {Dai}, {Reynolds}, \&
  {Kallman}}]{Kara2018}
{Kara}, E., {Dai}, L., {Reynolds}, C.~S., \& {Kallman}, T. 2018, \mnras, 474,
  3593, \dodoi{10.1093/mnras/stx3004}

\bibitem[{{King} \& {Pounds}(2015)}]{King2015}
{King}, A., \& {Pounds}, K. 2015, \araa, 53, 115,
  \dodoi{10.1146/annurev-astro-082214-122316}

\bibitem[{{Komossa} \& {Bade}(1999)}]{Komossa1999}
{Komossa}, S., \& {Bade}, N. 1999, \aap, 343, 775.
\newblock \doarXiv{astro-ph/9901141}

\bibitem[{{Komossa} {et~al.}(2004){Komossa}, {Halpern}, {Schartel}, {Hasinger},
  {Santos-Lleo}, \& {Predehl}}]{Komossa2004}
{Komossa}, S., {Halpern}, J., {Schartel}, N., {et~al.} 2004, \apjl, 603, L17,
  \dodoi{10.1086/382046}

\bibitem[{{Kubota} \& {Makishima}(2004)}]{Kubota2004}
{Kubota}, A., \& {Makishima}, K. 2004, \apj, 601, 428, \dodoi{10.1086/380433}

\bibitem[{{Leahy} {et~al.}(1983){Leahy}, {Darbro}, {Elsner}, {Weisskopf},
  {Sutherland}, {Kahn}, \& {Grindlay}}]{Leahy1983}
{Leahy}, D.~A., {Darbro}, W., {Elsner}, R.~F., {et~al.} 1983, \apj, 266, 160,
  \dodoi{10.1086/160766}

\bibitem[{{Lin} {et~al.}(2015){Lin}, {Maksym}, {Irwin}, {Komossa}, {Webb},
  {Godet}, {Barret}, {Grupe}, \& {Gwyn}}]{Lin2015}
{Lin}, D., {Maksym}, P.~W., {Irwin}, J.~A., {et~al.} 2015, \apj, 811, 43,
  \dodoi{10.1088/0004-637X/811/1/43}

\bibitem[{{Lipunova}(1999)}]{Lipunova1999}
{Lipunova}, G.~V. 1999, Astronomy Letters, 25, 508.
\newblock \doarXiv{astro-ph/9906324}

\bibitem[{{Liu} {et~al.}(2023){Liu}, {Malyali}, {Krumpe}, {Homan}, {Goodwin},
  {Grotova}, {Kawka}, {Rau}, {Merloni}, {Anderson}, {Miller-Jones},
  {Markowitz}, {Ciroi}, {Di Mille}, {Schramm}, {Tang}, {Buckley}, {Gromadzki},
  {Jin}, \& {Buchner}}]{Liu2023}
{Liu}, Z., {Malyali}, A., {Krumpe}, M., {et~al.} 2023, \aap, 669, A75,
  \dodoi{10.1051/0004-6361/202244805}

\bibitem[{{Lyubarskii}(1997)}]{Lyubarskii1997}
{Lyubarskii}, Y.~E. 1997, \mnras, 292, 679, \dodoi{10.1093/mnras/292.3.679}

\bibitem[{{Maccarone}(2003)}]{Maccarone2003}
{Maccarone}, T.~J. 2003, \aap, 409, 697, \dodoi{10.1051/0004-6361:20031146}

\bibitem[{{Malyali} {et~al.}(2021){Malyali}, {Rau}, {Merloni}, {Nandra},
  {Buchner}, {Liu}, {Gezari}, {Sollerman}, {Shappee}, {Trakhtenbrot}, {Arcavi},
  {Ricci}, {van Velzen}, {Goobar}, {Frederick}, {Kawka}, {Tartaglia}, {Burke},
  {Hiramatsu}, {Schramm}, {van der Boom}, {Anderson}, {Miller-Jones}, {Bellm},
  {Drake}, {Duev}, {Fremling}, {Graham}, {Masci}, {Rusholme}, {Soumagnac}, \&
  {Walters}}]{Malyali2021}
{Malyali}, A., {Rau}, A., {Merloni}, A., {et~al.} 2021, \aap, 647, A9,
  \dodoi{10.1051/0004-6361/202039681}

\bibitem[{{Markowitz}(2010)}]{Markowitz2010}
{Markowitz}, A. 2010, \apj, 724, 26, \dodoi{10.1088/0004-637X/724/1/26}

\bibitem[{Middleton \& Done(2010)}]{Middleton2010}
Middleton, M., \& Done, C. 2010, Monthly Notices of the Royal Astronomical
  Society, 403, 9, \dodoi{10.1111/j.1365-2966.2009.15969.x}

\bibitem[{{Middleton} {et~al.}(2015{\natexlab{a}}){Middleton}, {Heil},
  {Pintore}, {Walton}, \& {Roberts}}]{Middleton2015}
{Middleton}, M.~J., {Heil}, L., {Pintore}, F., {Walton}, D.~J., \& {Roberts},
  T.~P. 2015{\natexlab{a}}, \mnras, 447, 3243, \dodoi{10.1093/mnras/stu2644}

\bibitem[{{Middleton} {et~al.}(2011){Middleton}, {Sutton}, \&
  {Roberts}}]{Middleton2011}
{Middleton}, M.~J., {Sutton}, A.~D., \& {Roberts}, T.~P. 2011, \mnras, 417,
  464, \dodoi{10.1111/j.1365-2966.2011.19285.x}

\bibitem[{{Middleton} {et~al.}(2015{\natexlab{b}}){Middleton}, {Walton},
  {Fabian}, {Roberts}, {Heil}, {Pinto}, {Anderson}, \&
  {Sutton}}]{Middleton2015b}
{Middleton}, M.~J., {Walton}, D.~J., {Fabian}, A., {et~al.} 2015{\natexlab{b}},
  \mnras, 454, 3134, \dodoi{10.1093/mnras/stv2214}

\bibitem[{{Miller} {et~al.}(2015){Miller}, {Kaastra}, {Miller}, {Reynolds},
  {Brown}, {Cenko}, {Drake}, {Gezari}, {Guillochon}, {Gultekin}, {Irwin},
  {Levan}, {Maitra}, {Maksym}, {Mushotzky}, {O'Brien}, {Paerels}, {de Plaa},
  {Ramirez-Ruiz}, {Strohmayer}, \& {Tanvir}}]{Miller2015}
{Miller}, J.~M., {Kaastra}, J.~S., {Miller}, M.~C., {et~al.} 2015, \nat, 526,
  542, \dodoi{10.1038/nature15708}

\bibitem[{{Motta} {et~al.}(2021){Motta}, {Kajava}, {Giustini}, {Williams}, {Del
  Santo}, {Fender}, {Green}, {Heywood}, {Rhodes}, {Segreto}, {Sivakoff}, \&
  {Woudt}}]{Motta2021}
{Motta}, S.~E., {Kajava}, J.~J.~E., {Giustini}, M., {et~al.} 2021, \mnras, 503,
  152, \dodoi{10.1093/mnras/stab511}

\bibitem[{{Mu{\~n}oz-Darias} {et~al.}(2011){Mu{\~n}oz-Darias}, {Motta}, \&
  {Belloni}}]{Munoz2011}
{Mu{\~n}oz-Darias}, T., {Motta}, S., \& {Belloni}, T.~M. 2011, \mnras, 410,
  679, \dodoi{10.1111/j.1365-2966.2010.17476.x}

\bibitem[{{Mucciarelli} {et~al.}(2006){Mucciarelli}, {Casella}, {Belloni},
  {Zampieri}, \& {Ranalli}}]{Mucciarelli2006}
{Mucciarelli}, P., {Casella}, P., {Belloni}, T., {Zampieri}, L., \& {Ranalli},
  P. 2006, \mnras, 365, 1123, \dodoi{10.1111/j.1365-2966.2005.09754.x}

\bibitem[{{Mummery}(2021)}]{Mummery2021}
{Mummery}, A. 2021, \mnras, 507, L24, \dodoi{10.1093/mnrasl/slab088}

\bibitem[{{Narayan} {et~al.}(2017){Narayan}, {Sa{\`I}{\textsection}dowski}, \&
  {Soria}}]{Narayan2017}
{Narayan}, R., {Sa{\`I}{\textsection}dowski}, A., \& {Soria}, R. 2017, \mnras,
  469, 2997, \dodoi{10.1093/mnras/stx1027}

\bibitem[{{Narayan} \& {Yi}(1994)}]{Narayan1994}
{Narayan}, R., \& {Yi}, I. 1994, \apjl, 428, L13, \dodoi{10.1086/187381}

\bibitem[{{Neilsen} {et~al.}(2012){Neilsen}, {Remillard}, \&
  {Lee}}]{Neilsen2012}
{Neilsen}, J., {Remillard}, R.~A., \& {Lee}, J.~C. 2012, \apj, 750, 71,
  \dodoi{10.1088/0004-637X/750/1/71}

\bibitem[{{Noda} \& {Done}(2018)}]{Noda2018}
{Noda}, H., \& {Done}, C. 2018, \mnras, 480, 3898,
  \dodoi{10.1093/mnras/sty2032}

\bibitem[{{Papadakis} \& {Lawrence}(1993)}]{Papadakis1993}
{Papadakis}, I.~E., \& {Lawrence}, A. 1993, \mnras, 261, 612,
  \dodoi{10.1093/mnras/261.3.612}

\bibitem[{{Pasham} {et~al.}(2019){Pasham}, {Remillard}, {Fragile}, {Franchini},
  {Stone}, {Lodato}, {Homan}, {Chakrabarty}, {Baganoff}, {Steiner}, {Coughlin},
  \& {Pasham}}]{Pasham2019}
{Pasham}, D.~R., {Remillard}, R.~A., {Fragile}, P.~C., {et~al.} 2019, Science,
  363, 531, \dodoi{10.1126/science.aar7480}

\bibitem[{{Pasham} {et~al.}(2021){Pasham}, {Ho}, {Alston}, {Remillard}, {Ng},
  {Gendreau}, {Metzger}, {Altamirano}, {Chakrabarty}, {Fabian}, {Miller},
  {Bult}, {Arzoumanian}, {Steiner}, {Strohmayer}, {Tombesi}, {Homan},
  {Cackett}, \& {Harding}}]{Pasham2021}
{Pasham}, D.~R., {Ho}, W. C.~G., {Alston}, W., {et~al.} 2021, arXiv e-prints,
  arXiv:2112.04531.
\newblock \doarXiv{2112.04531}

\bibitem[{{Pasham} {et~al.}(2022){Pasham}, {Lucchini}, {Laskar}, {Gompertz},
  {Srivastav}, {Nicholl}, {Smartt}, {Miller-Jones}, {Alexander}, {Fender},
  {Smith}, {Fulton}, {Dewangan}, {Gendreau}, {Coughlin}, {Rhodes}, {Horesh},
  {van Velzen}, {Sfaradi}, {Guolo}, {Castro Segura}, {Aamer}, {Anderson},
  {Arcavi}, {Brennan}, {Chambers}, {Charalampopoulos}, {Chen}, {Clocchiatti},
  {de Boer}, {Dennefeld}, {Ferrara}, {Galbany}, {Gao}, {Gillanders}, {Goodwin},
  {Gromadzki}, {Huber}, {Jonker}, {Joshi}, {Kara}, {Killestein}, {Kosec},
  {Kocevski}, {Leloudas}, {Lin}, {Margutti}, {Mattila}, {Moore},
  {Muller-Bravo}, {Ngeow}, {Oates}, {Onori}, {Pan}, {Perez-Torres}, {Rani},
  {Remillard}, {Ridley}, {Schulze}, {Sheng}, {Shingles}, {Smith}, {Steiner},
  {Wainscoat}, {Wevers}, \& {Yang}}]{Pasham2022}
{Pasham}, D.~R., {Lucchini}, M., {Laskar}, T., {et~al.} 2022, arXiv e-prints,
  arXiv:2211.16537.
\newblock \doarXiv{2211.16537}

\bibitem[{{Petrucci} {et~al.}(2018){Petrucci}, {Ursini}, {De Rosa}, {Bianchi},
  {Cappi}, {Matt}, {Dadina}, \& {Malzac}}]{Petrucci2018}
{Petrucci}, P.~O., {Ursini}, F., {De Rosa}, A., {et~al.} 2018, \aap, 611, A59,
  \dodoi{10.1051/0004-6361/201731580}

\bibitem[{{Pinto} {et~al.}(2016){Pinto}, {Middleton}, \& {Fabian}}]{Pinto2016}
{Pinto}, C., {Middleton}, M.~J., \& {Fabian}, A.~C. 2016, \nat, 533, 64,
  \dodoi{10.1038/nature17417}

\bibitem[{{Rees}(1984)}]{Rees1984}
{Rees}, M.~J. 1984, \araa, 22, 471, \dodoi{10.1146/annurev.aa.22.090184.002351}

\bibitem[{{Reis} {et~al.}(2012){Reis}, {Miller}, {Reynolds}, {G{\"u}ltekin},
  {Maitra}, {King}, \& {Strohmayer}}]{Reis2012}
{Reis}, R.~C., {Miller}, J.~M., {Reynolds}, M.~T., {et~al.} 2012, Science, 337,
  949, \dodoi{10.1126/science.1223940}

\bibitem[{{Remillard} \& {McClintock}(2006)}]{Remillard2006}
{Remillard}, R.~A., \& {McClintock}, J.~E. 2006, \araa, 44, 49,
  \dodoi{10.1146/annurev.astro.44.051905.092532}

\bibitem[{{Remillard} {et~al.}(2021){Remillard}, {Loewenstein}, {Steiner},
  {Prigozhin}, {LaMarr}, {Enoto}, {Gendreau}, {Arzoumanian}, {Markwardt},
  {Basak}, {Stevens}, {Ray}, {Altamirano}, \& {Buisson}}]{Remillard2021}
{Remillard}, R.~A., {Loewenstein}, M., {Steiner}, J.~F., {et~al.} 2021, arXiv
  e-prints, arXiv:2105.09901.
\newblock \doarXiv{2105.09901}

\bibitem[{{Ross} \& {Fabian}(1993)}]{Ross1993}
{Ross}, R.~R., \& {Fabian}, A.~C. 1993, \mnras, 261, 74,
  \dodoi{10.1093/mnras/261.1.74}

\bibitem[{{Sadowski} \& {Narayan}(2016)}]{Sadowski2016}
{Sadowski}, A., \& {Narayan}, R. 2016, \mnras, 456, 3929,
  \dodoi{10.1093/mnras/stv2941}

\bibitem[{{Saxton} {et~al.}(2012{\natexlab{a}}){Saxton}, {Soria}, {Wu}, \&
  {Kuin}}]{Saxton2012b}
{Saxton}, C.~J., {Soria}, R., {Wu}, K., \& {Kuin}, N. P.~M. 2012{\natexlab{a}},
  \mnras, 422, 1625, \dodoi{10.1111/j.1365-2966.2012.20739.x}

\bibitem[{{Saxton} {et~al.}(2021){Saxton}, {Komossa}, {Auchettl}, \&
  {Jonker}}]{Saxton2021}
{Saxton}, R., {Komossa}, S., {Auchettl}, K., \& {Jonker}, P.~G. 2021, \ssr,
  217, 18, \dodoi{10.1007/s11214-020-00759-7}

\bibitem[{{Saxton} {et~al.}(2012{\natexlab{b}}){Saxton}, {Read}, {Esquej},
  {Komossa}, {Dougherty}, {Rodriguez-Pascual}, \& {Barrado}}]{Saxton2012a}
{Saxton}, R.~D., {Read}, A.~M., {Esquej}, P., {et~al.} 2012{\natexlab{b}},
  \aap, 541, A106, \dodoi{10.1051/0004-6361/201118367}

\bibitem[{{Scaringi} {et~al.}(2012){Scaringi}, {K{\"o}rding}, {Uttley},
  {Knigge}, {Groot}, \& {Still}}]{Scaringi2012}
{Scaringi}, S., {K{\"o}rding}, E., {Uttley}, P., {et~al.} 2012, \mnras, 421,
  2854, \dodoi{10.1111/j.1365-2966.2012.20512.x}

\bibitem[{{Shakura} \& {Sunyaev}(1973)}]{Shakura1973}
{Shakura}, N.~I., \& {Sunyaev}, R.~A. 1973, \aap, 500, 33

\bibitem[{{Singh} {et~al.}(1985){Singh}, {Garmire}, \& {Nousek}}]{Singh1985}
{Singh}, K.~P., {Garmire}, G.~P., \& {Nousek}, J. 1985, \apj, 297, 633,
  \dodoi{10.1086/163560}

\bibitem[{{Sobolewska} {et~al.}(2011){Sobolewska}, {Papadakis}, {Done}, \&
  {Malzac}}]{Sobolewska2011}
{Sobolewska}, M.~A., {Papadakis}, I.~E., {Done}, C., \& {Malzac}, J. 2011,
  \mnras, 417, 280, \dodoi{10.1111/j.1365-2966.2011.19209.x}

\bibitem[{{Stiele} \& {Kong}(2017)}]{Stiele2017}
{Stiele}, H., \& {Kong}, A.~K.~H. 2017, \apj, 844, 8,
  \dodoi{10.3847/1538-4357/aa774e}

\bibitem[{{Strohmayer} \& {Mushotzky}(2003)}]{Strohmayer2003}
{Strohmayer}, T.~E., \& {Mushotzky}, R.~F. 2003, \apjl, 586, L61,
  \dodoi{10.1086/374732}

\bibitem[{{Sutton} {et~al.}(2013){Sutton}, {Roberts}, \&
  {Middleton}}]{Sutton2013}
{Sutton}, A.~D., {Roberts}, T.~P., \& {Middleton}, M.~J. 2013, \mnras, 435,
  1758, \dodoi{10.1093/mnras/stt1419}

\bibitem[{{Takeuchi} {et~al.}(2013){Takeuchi}, {Ohsuga}, \&
  {Mineshige}}]{Takeuchi2013}
{Takeuchi}, S., {Ohsuga}, K., \& {Mineshige}, S. 2013, \pasj, 65, 88,
  \dodoi{10.1093/pasj/65.4.88}

\bibitem[{{Takeuchi} {et~al.}(2014){Takeuchi}, {Ohsuga}, \&
  {Mineshige}}]{Takeuchi2014}
---. 2014, \pasj, 66, 48, \dodoi{10.1093/pasj/psu011}

\bibitem[{{Uttley} \& {McHardy}(2001)}]{Uttley2001}
{Uttley}, P., \& {McHardy}, I.~M. 2001, \mnras, 323, L26,
  \dodoi{10.1046/j.1365-8711.2001.04496.x}

\bibitem[{{Uttley} {et~al.}(2002){Uttley}, {McHardy}, \&
  {Papadakis}}]{Uttley2002}
{Uttley}, P., {McHardy}, I.~M., \& {Papadakis}, I.~E. 2002, \mnras, 332, 231,
  \dodoi{10.1046/j.1365-8711.2002.05298.x}

\bibitem[{{van Velzen} {et~al.}(2020){van Velzen}, {Holoien}, {Onori}, {Hung},
  \& {Arcavi}}]{Velzen2020}
{van Velzen}, S., {Holoien}, T. W.~S., {Onori}, F., {Hung}, T., \& {Arcavi}, I.
  2020, \ssr, 216, 124, \dodoi{10.1007/s11214-020-00753-z}

\bibitem[{{van Velzen} {et~al.}(2021){van Velzen}, {Gezari}, {Hammerstein},
  {Roth}, {Frederick}, {Ward}, {Hung}, {Cenko}, {Stein}, {Perley}, {Taggart},
  {Foley}, {Sollerman}, {Blagorodnova}, {Andreoni}, {Bellm}, {Brinnel}, {De},
  {Dekany}, {Feeney}, {Fremling}, {Giomi}, {Golkhou}, {Graham}, {Ho},
  {Kasliwal}, {Kilpatrick}, {Kulkarni}, {Kupfer}, {Laher}, {Mahabal}, {Masci},
  {Miller}, {Nordin}, {Riddle}, {Rusholme}, {van Santen}, {Sharma}, {Shupe}, \&
  {Soumagnac}}]{van_Velzen_21}
{van Velzen}, S., {Gezari}, S., {Hammerstein}, E., {et~al.} 2021, \apj, 908, 4,
  \dodoi{10.3847/1538-4357/abc258}

\bibitem[{{VanderPlas}(2018)}]{VanderPlas2018}
{VanderPlas}, J.~T. 2018, \apjs, 236, 16, \dodoi{10.3847/1538-4365/aab766}

\bibitem[{{Vaughan}(2005)}]{Vaughan2005}
{Vaughan}, S. 2005, \aap, 431, 391, \dodoi{10.1051/0004-6361:20041453}

\bibitem[{{Vaughan} {et~al.}(2003){Vaughan}, {Edelson}, {Warwick}, \&
  {Uttley}}]{Vaughan2003}
{Vaughan}, S., {Edelson}, R., {Warwick}, R.~S., \& {Uttley}, P. 2003, \mnras,
  345, 1271, \dodoi{10.1046/j.1365-2966.2003.07042.x}

\bibitem[{{Wang} {et~al.}(2018){Wang}, {M{\'e}ndez}, {Altamirano}, {Court},
  {Beri}, \& {Cheng}}]{Wang2018}
{Wang}, Y., {M{\'e}ndez}, M., {Altamirano}, D., {et~al.} 2018, \mnras, 478,
  4837, \dodoi{10.1093/mnras/sty1372}

\bibitem[{{Wang} {et~al.}(2020){Wang}, {Ji}, {Zhang}, {M{\'e}ndez}, {Qu},
  {Maggi}, {Ge}, {Qiao}, {Tao}, {Zhang}, {Altamirano}, {Zhang}, {Ma}, {Lu},
  {Li}, {Huang}, {Zheng}, {Chen}, {Chang}, {Tuo}, {G{\"u}ng{\"o}r}, {Song},
  {Xu}, {Cao}, {Chen}, {Liu}, {Bu}, {Cai}, {Chen}, {Chen}, {Chen}, {Chen},
  {Cui}, {Cui}, {Deng}, {Dong}, {Du}, {Fu}, {Gao}, {Gao}, {Gao}, {Gu}, {Guan},
  {Guo}, {Han}, {Huo}, {Jia}, {Jiang}, {Jiang}, {Jin}, {Jin}, {Kong}, {Li},
  {Li}, {Li}, {Li}, {Li}, {Li}, {Li}, {Li}, {Li}, {Li}, {Liang}, {Liao}, {Liu},
  {Liu}, {Liu}, {Liu}, {Lu}, {Lu}, {Luo}, {Luo}, {Meng}, {Nang}, {Nie}, {Ou},
  {Sai}, {Shang}, {Song}, {Sun}, {Tan}, {Wang}, {Wang}, {Wang}, {Wang}, {Wang},
  {Wen}, {Wu}, {Wu}, {Wu}, {Xiao}, {Xiao}, {Xiong}, {Yang}, {Yang}, {Yang},
  {Yang}, {Yi}, {Yin}, {You}, {Zhang}, {Zhang}, {Zhang}, {Zhang}, {Zhang},
  {Zhang}, {Zhang}, {Zhang}, {Zhang}, {Zhang}, {Zhang}, {Zhang}, {Zhang},
  {Zhang}, {Zhang}, {Zhang}, {Zhao}, {Zhao}, {Zhou}, {Zhou}, {Zhuang}, {Zhu},
  {Zhu}, \& {Wang}}]{Wang2020}
{Wang}, Y., {Ji}, L., {Zhang}, S.~N., {et~al.} 2020, \apj, 896, 33,
  \dodoi{10.3847/1538-4357/ab8db4}

\bibitem[{{Wang} {et~al.}(2023){Wang}, {Baldi}, {del Palacio}, {Guolo}, {Yang},
  {Zhang}, {Done}, {Castro Segura}, {Pasham}, {Middleton}, {Altamirano},
  {Gandhi}, {Qiao}, {Jiang}, {Yan}, {Giroletti}, {Migliori}, {McHardy},
  {Panessa}, {Jin}, {Shen}, \& {Dai}}]{Wang2023}
{Wang}, Y., {Baldi}, R.~D., {del Palacio}, S., {et~al.} 2023, \mnras, 520,
  2417, \dodoi{10.1093/mnras/stad101}

\bibitem[{{Wevers} {et~al.}(2021){Wevers}, {Pasham}, {van Velzen},
  {Miller-Jones}, {Uttley}, {Gendreau}, {Remillard}, {Arzoumanian},
  {L{\"o}wenstein}, \& {Chiti}}]{Wevers2021}
{Wevers}, T., {Pasham}, D.~R., {van Velzen}, S., {et~al.} 2021, \apj, 912, 151,
  \dodoi{10.3847/1538-4357/abf5e2}

\bibitem[{{Wevers} {et~al.}(2023){Wevers}, {Coughlin}, {Pasham}, {Guolo},
  {Sun}, {Wen}, {Jonker}, {Zabludoff}, {Malyali}, {Arcodia}, {Liu}, {Merloni},
  {Rau}, {Grotova}, {Short}, \& {Cao}}]{Wevers2023}
{Wevers}, T., {Coughlin}, E.~R., {Pasham}, D.~R., {et~al.} 2023, \apjl, 942,
  L33, \dodoi{10.3847/2041-8213/ac9f36}

\bibitem[{{White} \& {Marshall}(1984)}]{White1984}
{White}, N.~E., \& {Marshall}, F.~E. 1984, \apj, 281, 354,
  \dodoi{10.1086/162104}

\bibitem[{{Wolff} {et~al.}(2021){Wolff}, {Guillot}, {Bogdanov}, {Ray}, {Kerr},
  {Arzoumanian}, {Gendreau}, {Miller}, {Dittmann}, {Ho}, {Guillemot},
  {Cognard}, {Theureau}, \& {Wood}}]{Wolff2021}
{Wolff}, M.~T., {Guillot}, S., {Bogdanov}, S., {et~al.} 2021, \apjl, 918, L26,
  \dodoi{10.3847/2041-8213/ac158e}

\bibitem[{{Wu} {et~al.}(2016){Wu}, {Czerny}, {Grzedzielski}, {Janiuk}, {Gu},
  {Dong}, {Cao}, {You}, {Yan}, \& {Sun}}]{Wu2016}
{Wu}, Q., {Czerny}, B., {Grzedzielski}, M., {et~al.} 2016, \apj, 833, 79,
  \dodoi{10.3847/1538-4357/833/1/79}

\bibitem[{{Yao} {et~al.}(2022){Yao}, {Lu}, {Guolo}, {Pasham}, {Gezari},
  {Gilfanov}, {Gendreau}, {Harrison}, {Cenko}, {Kulkarni}, {Miller}, {Walton},
  {Garc{\'\i}a}, {van Velzen}, {Alexander}, {Miller-Jones}, {Nicholl},
  {Hammerstein}, {Medvedev}, {Stern}, {Ravi}, {Sunyaev}, {Bloom}, {Graham},
  {Kool}, {Mahabal}, {Masci}, {Purdum}, {Rusholme}, {Sharma}, {Smith}, \&
  {Sollerman}}]{yao2022}
{Yao}, Y., {Lu}, W., {Guolo}, M., {et~al.} 2022, \apj, 937, 8,
  \dodoi{10.3847/1538-4357/ac898a}

\bibitem[{{Yao} {et~al.}(2023){Yao}, {Ravi}, {Gezari}, {van Velzen}, {Lu},
  {Schulze}, {Somalwar}, {Kulkarni}, {Hammerstein}, {Nicholl}, {Graham},
  {Perley}, {Cenko}, {Stein}, {Ricarte}, {Chadayammuri}, {Quataert}, {Bellm},
  {Bloom}, {Dekany}, {Drake}, {Groom}, {Mahabal}, {Prince}, {Riddle},
  {Rusholme}, {Sharma}, {Sollerman}, \& {Yan}}]{Yao2023}
{Yao}, Y., {Ravi}, V., {Gezari}, S., {et~al.} 2023, arXiv e-prints,
  arXiv:2303.06523, \dodoi{10.48550/arXiv.2303.06523}

\bibitem[{{Zauderer} {et~al.}(2013){Zauderer}, {Berger}, {Margutti}, {Pooley},
  {Sari}, {Soderberg}, {Brunthaler}, \& {Bietenholz}}]{Zauderer2013}
{Zauderer}, B.~A., {Berger}, E., {Margutti}, R., {et~al.} 2013, \apj, 767, 152,
  \dodoi{10.1088/0004-637X/767/2/152}

\end{thebibliography}
\bibliographystyle{aasjournal}

\end{document}